\documentclass[letterpaper]{article} 
\usepackage{aaai2026}  
\nocopyright
\usepackage{times}  
\usepackage{helvet}  
\usepackage{courier}  
\usepackage[hyphens]{url}  
\usepackage{graphicx} 

\usepackage{booktabs}
\usepackage{tabularx,array,ragged2e,threeparttable}
\usepackage{amssymb}
\newcolumntype{L}[1]{%
  >{\hsize=#1\hsize\linewidth=\hsize
    \RaggedRight
    \hyphenpenalty=10000
    \exhyphenpenalty=10000
    \arraybackslash}X%
}

\usepackage{natbib}  
\usepackage{xcolor}
\usepackage{caption} 
\usepackage{algorithm}
\usepackage{algorithmic}
\usepackage{amssymb}
\usepackage{pifont}

\usepackage{tabularx,array,multirow,pifont}
\usepackage{colortbl}

\definecolor{aigcyes}{HTML}{E1EFD9}
\definecolor{aigcdesc}{HTML}{FFF2CC}
\definecolor{aigcno}{HTML}{F5E3D8}
\definecolor{aigcna}{HTML}{F2F2F2}

\usepackage{tikz}
\usetikzlibrary{patterns}

\DeclareRobustCommand{\solidkey}{%
  \ensuremath{\vcenter{\hbox{%
    \tikz \draw[
      draw=black!65, fill=gray!15, line width=0.4pt
    ] (0,0) rectangle (0.9em,0.65em);%
  }}}%
}

\DeclareRobustCommand{\stripedkey}{%
  \ensuremath{\vcenter{\hbox{%
    \tikz \draw[
      draw=black!65, fill=gray!15,
      pattern=north east lines,
      pattern color=black!65, line width=0.4pt
    ] (0,0) rectangle (0.9em,0.65em);%
  }}}%
}

\usepackage{newfloat}
\usepackage{listings}
\DeclareCaptionStyle{ruled}{labelfont=normalfont,labelsep=colon,strut=off} 
\floatstyle{ruled}
\newfloat{listing}{tb}{lst}{}
\floatname{listing}{Listing}
\title{Drowning in \textit{AI Slop}: How Social Media Platforms (Do Not) Label AI and Deepfake Content under EU law} 
\author {
    Bram Rijsbosch\textsuperscript{\rm 1},
    Luka Bekavac\textsuperscript{\rm 2},
    Henry Tari\textsuperscript{\rm 1},
    Gijs van Dijck\textsuperscript{\rm 1},
    Konrad Kollnig\textsuperscript{\rm 1}
}
\affiliations {
    \textsuperscript{\rm 1}Maastricht University, Law \& Tech Lab\\
    \textsuperscript{\rm 2}University St. Gallen, Institute of Computer Science\\
}

\begin{document}

\maketitle

\begin{abstract}
AI labels are emerging as a primary safeguard for transparency about AI-generated content on social media, including under the EU Digital Services Act and AI Act. Yet, limited systematic evidence exists on how platforms implement such labels in practice. We conduct a legally grounded audit of AI labelling across Instagram, TikTok, X, and YouTube, drawing on the European Commission's July 2026 guidelines on deepfakes.
To this end, we analyse platform policies and detection approaches, 10,722 posts collected via systemic-risk and AI-related keywords, an expert-annotated subset of 500 posts, and controlled uploads to the four platforms of outputs from ten popular generative AI tools.
We find that AI labelling is now broadly established. All four platforms apply labels automatically and a greater share of labels are platform-applied than in earlier audits. However, coverage remains incomplete where labelling arguably matters most: only 33\% of expert-identified deepfakes in systemic risk contexts carried a platform-applied AI label, while reaching a median of 160,000 views. In controlled uploads of AI-generated content carrying standard AI provenance signals, platforms labelled only 61\% of uploads, and commonly strip those signals after uploading. Overall, platform rules, label designs, detection approaches, and reporting diverge substantially. In response, we identify concrete opportunities to improve the uptake, clarity and efficacy of AI labelling.\\\textbf{Note: This work is under submission and has not undergone peer review yet.}
\end{abstract}

\section{Introduction}
\begin{figure}
    \centering
    \includegraphics[width=\columnwidth]{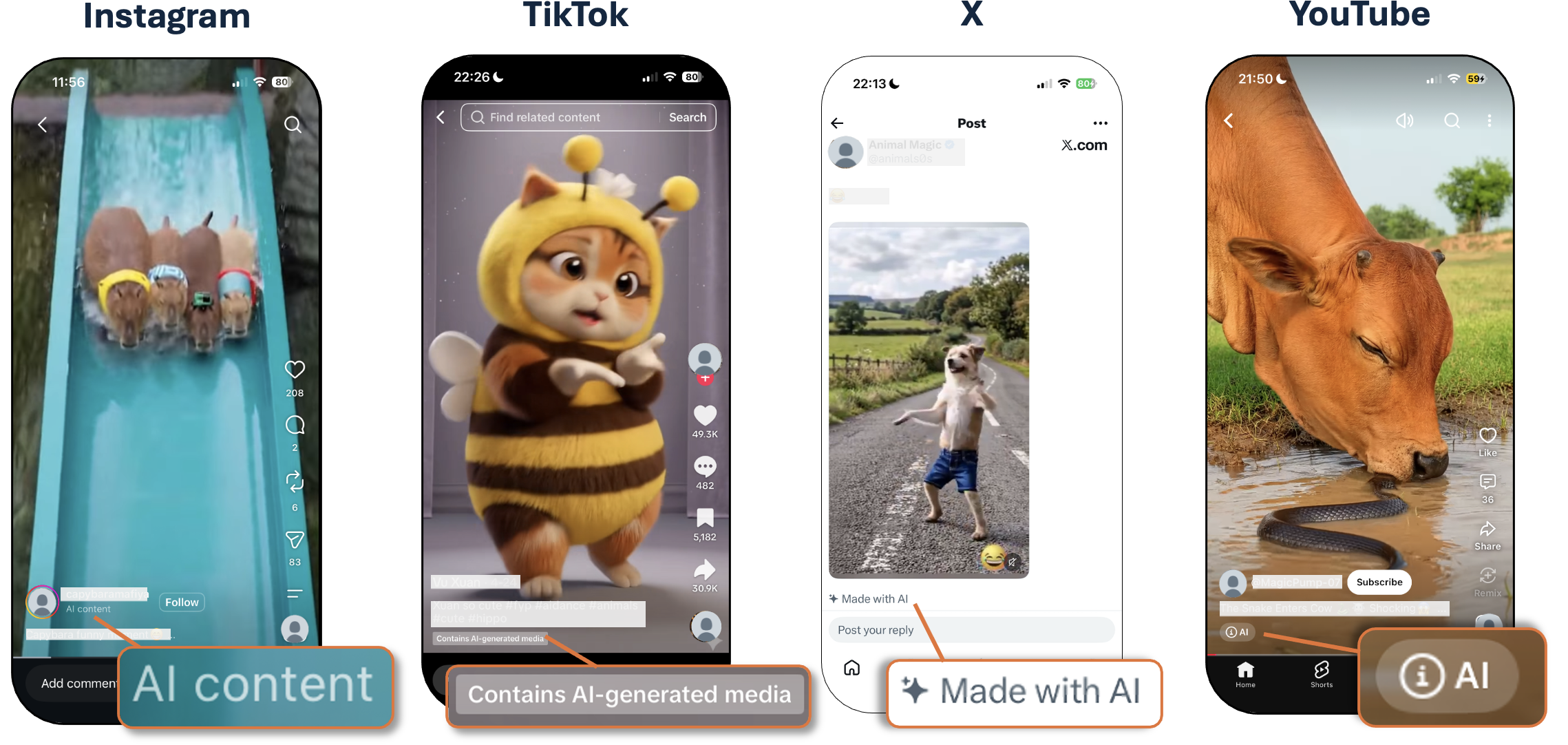}
    \caption{Examples of AI labels on Instagram, TikTok, X, and YouTube. Screenshots from public posts on iOS interface, taken on 13 August 2026 (Instagram: 20 August). Appendix~\ref{appendix_all_labels} provides an overview of all AI label types found.}
    \label{fig:labels}
\end{figure}

Social media platforms are increasingly flooded with AI-generated content (AIGC) that can be produced at low cost and great speed \cite{aiforensics-core, aiforensics-agentic-accounts}, sometimes also referred to as \textit{AI slop}. 
As AIGC becomes more widespread and more realistic, regulators, civil society and social media platforms themselves warn of the growing risks to civic discourse, electoral integrity, and public trust \cite{eu_board_systemicrisks, WEF2025GlobalRisks, meta_systemicrisk_2025}.  
In response, AI labels --- user-facing indicators highlighting that content has been AI-generated or manipulated (see Figure~\ref{fig:labels}) --- have emerged, intended to help users recognise AIGC and support platform moderation and curation. Most of the largest social media platforms have adopted AI-labelling measures in recent years \cite{gao2026governance}.

Measures for marking and disclosing AIGC have since also become a legal obligation in several major jurisdictions, including the EU, China, and California \cite{eu_aia_2024,CAC2025AILabeling, CaliforniaSB942}, while others are considering doing so \cite{Microsoft2026MediaIntegrity}.
Particularly in the EU, Article 50 of the AI Act (as of August 2026) requires providers of generative AI systems to apply machine-readable markings to AIGC, and deployers of such systems to additionally apply visible disclosures to AI \textit{deepfakes} specifically \cite{eu_aia_2024}.
However, because generation-stage markings can easily be removed, social media platforms that spread AIGC also carry significant social and legal responsibility to act on such content. 
The main EU legal instrument for this responsibility is the 2022 Digital Services Act (DSA), which requires Very Large Online Platforms (VLOPs) to identify, assess, and mitigate systemic risks, including those posted by AIGC. The DSA lists the use of prominent markings for generated or manipulated content that can falsely appear as authentic or truthful --- what we will refer to as \textit{deepfakes} (see Section~\ref{3-3}) --- as an important such systemic risk mitigation measure  \cite{eu_dsa_2022, eu_dsa_code_of_conduct_2025, eu_dsa_electoral_guidelines}.

Despite this growing attention, little systematic and independent empirical evidence yet exists on how AI-labelling measures are implemented in practice by social media platforms, and how these implementations align with emerging legal requirements, such as those in the EU. 
Platforms so far disclose only limited information about their labelling practices and the resulting outcomes (Section \ref{4}). Independent empirical studies have generally been platform- or topic-specific, based on pre-2026 data, and not explicitly tied to the EU legal requirements (Section \ref{2}).
Related work has examined the legal requirements (e.g., \cite{ labuz2025teleological}), and discussed the design and effects of AI labels on users (e.g., \cite{ epstein2023label}), but neither strands have focused on actual AI-labelling outcomes or the related platform's detection approaches (Section \ref{2}). Closing this gap requires a consideration of the relevant legal rules, as well as the technical methods to collect platforms' label data at scale.

To address this technical-legal gap, this paper conducts a legally grounded, cross-platform audit of AI-labelling practices and outcomes on Instagram (Reels), TikTok, X, and YouTube (Shorts): the four largest social media platforms in the EU from distinct providers that are classified as VLOPs under the DSA \cite{socialmediatransparancy}. Specifically, we address the following three research questions: 
\begin{itemize}
    \item \textbf{RQ1} What content does EU law expect to be labelled on popular social media platforms? (Section~\ref{3})
    \item \textbf{RQ2} How have popular social media platforms operationalised these requirements in their stated policies, label designs, public reporting, and detection approaches? (Section~\ref{4})
    \item \textbf{RQ3} What do popular social media platforms actually label in practice, and to what extent is high-risk \textit{deepfake} content (as prioritised by EU law) labelled? (Section ~\ref{5}) 
\end{itemize}
To answer RQ3, we construct a large-scale dataset of 10,722 recent public image and video posts from DSA-relevant systemic risk contexts, combined with an expert-annotated subset of 500 posts.

\textbf{Structure.} Section~\ref{2} discusses related work. Section~\ref{3} provides an overview of the EU's legal rules for AIGC-transparency, and operationalises what this means for platform AI-labelling (RQ1). Section~\ref{4} examines platform's policies, designs, reporting, and detection practices (RQ2). Section \ref{5} measures actual AI-labelling outcomes (RQ3). Finally, Section \ref{Limitations} states important limitations of this study, and Section~\ref{7} discusses our findings and their implications for platforms, users, legal frameworks, policy-makers, and external auditors.

\section{Related Work}\label{2}
\paragraph{Empirical audits of AI-labelling.} Research on how AIGC is labelled on social media has generally reported labelling outcomes to be partial and uneven across platforms. In a June 2025 audit of Instagram and TikTok, AI Forensics \cite{aiforensics-core} found that roughly 25\% of posts in top search results for common non-AI hashtags contained AIGC, with only about half of this content on TikTok, and under a quarter on Instagram, carrying an AI label (including AI hashtags in captions). The majority of these labels were creator-applied. They also document several notable visibility gaps, such as Instagram's AI labels not appearing on its web interface.    
An August 2025 follow-up study focused on agentic AI accounts on TikTok and found fewer than half of their posts labelled, again mostly via creator-labels (30\%) or caption disclosures (12\%), rather than by TikTok itself (1\%) \cite{aiforensics-agentic-accounts}. Related small-scale audits by AI Forensics in electoral contexts similarly found most AIGC posts unlabelled \cite{aiforensics-artificial-elections, aiforensics-artificial-elections-2, aiforensics-dutch-elections, aiforensics-polish-elections}. 

Other academic studies report similar patterns: \citet{kuiperslabeling} found that 52\% of AI-hashtagged YouTube Shorts and 38\% TikTok videos carried platform AI labels, again mostly creator-applied. \citet{chrysidis2026synthetic} analysed five years of X posts that contained community notes, and found that 11\% of these notes had AI-generated references, while only 1\% of the underlying posts contained AI mentions in their captions. Using algorithmic classifiers, they report that around half of classified AI-generated image posts contained an AI mention.
Finally, an early 2024 study by \citet{diresta2024spammers} found no platform-based AI labels across 120 popular Facebook pages with a high share of AIGC, likely because the study predated Meta's AI-labelling roll-out. 

Together, these studies suggest that AI-labelling has so far reached at best around half of AIGC posts, and has rested largely on creator self-disclosures. These outcomes are, however, difficult to compare or aggregate, as these studies differ in sampling strategies, time periods, platform coverage, and in how they define AIGC that should be labelled. Plus, none of these studies relies on recent 2026 data, focuses specifically on \textit{deepfakes}, or grounds its methodological choices explicitly in (EU) law, as our study does.

\paragraph{Platform Policies and Practices.} A second strand of related work has examined what platforms say and do, rather than focusing on the labelling outcomes. \citet{gao2026governance} analysed the AIGC governance policies of 40 popular platforms (April-June 2025), finding that two-thirds explicitly govern AIGC through six main mechanisms: labelling AIGC (18 of 40 platforms), moderation under existing policies, additional restrictions, monetisation limits, safety measures for integrated AI tools, and user-facing literacy resources. On labelling, they report wide variation in what must be disclosed and in label design and placement, and note that only a minority of platforms describe their techniques for identifying AIGC. 
A January 2026 UK Parliamentary briefing also considered the labelling measures of several major platforms \cite{uk-parliament}, and reports that Meta, TikTok and YouTube each have AIGC disclosure requirements alongside automatic labelling measures, while X lacked specific AI-labelling measures apart from its Community Notes. 
Complementing this, a 2026 study of Indicator used uploads of AIGC from Meta's, Google's and OpenAI's models to five major platforms (Instagram, LinkedIn, Pinterest, TikTok, and YouTube) to find that these frequently failed to mark such content automatically, with YouTube labelling roughly 50\%, TikTok about a third, and Instagram only 14\% of the uploads \cite{Indicator2026AILabeling}. \citet{sekwenz2026content} further considered platforms' reporting data in the DSA Transparency Database, and reports that moderation measures for content tagged as `synthetic media' focused largely on visibility restrictions, rather than labelling. 

We build on this literature by analysing for recent policy changes, and extending these experiments with broader model coverage and more detailed platform-specific results.

\paragraph{Legal analyses of AIGC transparency requirements in the EU.} Legal scholarship has further analysed the EU's rules for AIGC transparency, including critical reflections on their wording, scope and the risks of circumventions \cite{labuz2024deep, meding2025constitutes}. Most relevant to our analysis is the question where the treshold lies in the EU for content to require a marking or disclosure (i.e., when content counts as AI-generated, manipulated or as a deepfake) \cite{sekwenz2026content}.
This is particularly debated for `deep fakes', for which the literature reports a lack of consensus about the scope of its definition as used in EU law, and beyond \cite{labuz2024deep, labuz2025teleological, sekwenz2026content, meding2025constitutes}. \citet{labuz2025teleological} for example argues for a teleological and systemic reading, so that deepfakes are not limited only to directly identifiable persons, subjects or events, as a strict reading would suggest. 
These debates, however, predate the Europeans Commission's July 2026 guidelines on the AI Act's transparency requirements, which includes a clarification of its deepfake definition. We therefore rely on these guidelines in Section \ref{3} to address these treshold question. 

\paragraph{AI label design, effects, and perceptions.} A final strand of literature has studied the design, perceptions and effects of AI-labels, mainly through user studies. Several studies report that AI labels can help inform users about the AI-generated nature of content and can help in countering misinformation \cite{gamage2025labeling, holtervennhoff2026s, pawelczyk2026implied, epstein2023label}, and that users generally value AI labels and would prefer to see all AIGC on social media labelled \cite{holtervennhoff2026s}. 
Others, however, also question the effectiveness of AI labels, noting for example that labels do not negate the negative impacts of AIGC, can reduce belief in true claims made with AIGC, and can increase belief in non-labelled misleading AIGC \cite{feng2023examining, clark2026, holtervennhoff2026s, pawelczyk2026implied, bechmann2026transparency}. Design-focused work further documents confusion over the meaning of AI-labels \cite{gamage2025labeling,burrus2024unmasking}, and considers the tradeoffs between process- and harm-based labels  \cite{epstein2023label,wittenberg2024labeling}. Studies also report that engagement can decline for AI-labelled content \cite{seeger2026ai, carney2026made}, which could disincentivise disclosures. Explicit engagement with label design and effect questions, however, lies outside the scope of this work. Instead, we aim to complement this literature with current evidence on labelling practices and outcomes on platforms. 

\section{RQ1: The EU's Legal Rules for AIGC Transparency and Deepfakes}\label{3}
The EU's transparency obligations for AIGC stem from two recent regulations, which apply at different stages of the content lifecycle: 
The 2024 AI Act imposes AI transparency obligations at the point of \textit{generation}, while the 2022 DSA governs the downstream \textit{dissemination} and moderation of such content on online platforms \cite{eu_aia_2024, eu_dsa_2022}. Our study is primarily grounded in the DSA, but the AI Act's obligations are complimentary and its recent implementation guidelines can help interpret what content platforms are expected to label (Section~\ref{3-3}). 

\subsection{DSA: Transparency at Dissemination.}\label{3-1}
The DSA regulates the operation of online intermediary services, including social media platforms. Its most stringent obligations apply to Very Large Online Platforms (VLOPs) and Very Large Online Search Engines (VLOSEs) that have more than 45 million average monthly users in the EU, which includes the four platforms of this study. Central to these large players is the DSA's obligation to identify, assess and mitigate any systemic risks that may arise from the design, functioning and use of their services (Articles 34-35). Systemic risks cover illegal content, and actual or foreseeable negative effects to civic discourse, electoral process, public security, public health, or the fundamental rights of citizens. 
In their 2025 report on the most prominent systemic risks, the European Commission and DSA Board highlights how growing spread of (realistic) AIGC already contributes to increased risks in each of these areas \cite{eu_board_systemicrisks}. 

The DSA, however, does not mandate any specific risk mitigation solutions, but the legal text does list several measures that could be considered by platforms. One concerns the labelling of generated or manipulated content, and states:
\begin{quote}
    \textit{``ensuring that an item of information, whether it constitutes a generated or manipulated image, audio or video that appreciably resembles existing persons, objects, places or other entities or events and falsely appears to a person to be authentic or truthful \textbf{is distinguishable through prominent markings} when presented on their online interfaces, and, in addition, providing an easy to use functionality which enables recipients of the service to indicate such information.'' (Art. 35(1)(k))}
\end{quote}
We refer to such content as \textit{deepfakes}, following the similarity of this wording to the AI Act's deep fake definition (AI Act Article 2(60)), as discussed in Section~\ref{3-3}. ,

Although this measure is thus not a direct legal obligation, accompanying legal implementation guidance does reinforces it as a clear expectation for VLOPs: The 2024 DSA guidelines on systemic risks in electoral contexts recommend that platforms clearly label (or otherwise make distinguishable through prominent markings) content meeting the definition of Article 35(1)(k) (which they too refer to as ``deepfakes''), and ensure that such labels are retained when content is re-shared \cite{eu_dsa_electoral_guidelines}. 
The 2025 EU Code of Conduct on Disinformation, which serves as a benchmark for compliance with disinformation risks, further commits signatories (including Instagram, TikTok and YouTube) to maintain policies for countering prohibited manipulative practices that involve generative AI systems, including the warning of users and the proactive detection of such content \cite{eu_dsa_code_of_conduct_2025}. Platforms are thereby additionally expected to address illegal content that is non-compliant with other EU laws (Article 3(H)), which may include AIGC that violates the EU's copyright or data protection rules, and could also be related to content violating the disclosure requirements of the AI Act (although scholars argue that such a violation is unlikely to trigger any removal obligations for platforms \cite{feltes2026regulating}).
Finally, platforms must also enforce their own terms \& conditions (Article 14), which includes any imposed AIGC labelling rules for their users. 

Neither the DSA nor its (public) guidance, however, specifies what an AI label should look like, which content falls within the definition above, how platforms should handle mislabelling risks, or what automated detection methods can and should be used. Mitigation measures must only be reasonable, proportionate and effective given the current state of the art (Article 35(1)). These open-ended expectations, combined with the current absence of systematic independent evidence, form a central motivations for this study.

\subsection{AI Act: Transparency at Generation.}\label{3-2}
Two AI Act transparency requirements are relevant for this study. First, providers of generative AI systems must ensure that all generated outputs carry robust and interoperable machine-readable markings (Article 50(2), which can include metadata identifiers, invisible watermarks, or other provenance signals (Recital 133). This obligation primarily aims to support the downstream detection of such content, such as by platforms.
Second, deployers generating or manipulating AI image, audio or video content that constitutes a deep fake must visibly disclose its AI-generated origin (Article 50(4)). The AI Act thereby defines deep fakes as ``\textit{AI-generated or manipulated image, audio, or video content that resembles existing persons, objects, places, entities or events and would falsely appear to a person to be authentic or truthful}'' (Article 3(60)).

Both obligations generally apply from 2 August 2026 (with a 4-month grace period for systems already on the markt). Their impact on platform-level transparency nevertheless remains uncertain: machine-readable markings are often still easily removable and widespread adoption might prove difficult to enforce \cite{Microsoft2026MediaIntegrity, rijsbosch2025watermark}. The AI Act also exempts individuals acting in a purely personal and non-professional capacity (Article 2(10), which is especially relevant to the deployers' deep fake disclosure requirement, as this exception could apply to a large share of social media users who act in a personal capacity to generate and disseminate AIGC  \cite{AI-Act-guidelines}. Platform-side labelling measures therefore likely remain essential for transparency on social media.

\subsection{What content must be labelled? Interpreting the EU's rules on deepfakes}\label{3-3}
A central interpretative question for our audit is determining which content falls within the DSA's labelling expectations. 
As no DSA-specific guidance for this yet exists, we draw on the European Commission's July 2026 guidelines on the AI Act's AI transparency obligations (including its deepfake definition) to help in operationalising the DSA's expectations in criteria that can inform our subsequent analyses and annotation \cite{AI-Act-guidelines}. We consider this choice justified because the wording of AI Act's deepfake definition is highly similar to that of Article 35(1)(k) in the DSA, as also confirmed in the guidelines. Plus, both provisions aim to address similar policy objectives, such as combatting risks of misinformation, manipulation. 

On this basis, we use the guidelines to derive an operational reading of what platforms are expected to label, which is structured in two main steps: (1) what counts as (AI) generated or manipulated, (2) what counts as a deepfake. We then use this operationalisation as a guiding benchmark to compare against platform's policies (Section~\ref{4}) and to structure the annotation codebook as used in Section~\ref{5}. 

\paragraph{Step 1: What counts as (AI) generated or manipulated?\\} Before the deepfake criteria can be applied, it must first be determined if content qualifies as generated or manipulated. The guidelines, however, offer little direct guidance here, as they are written primarily from the perspective of a deployer who already knows they are generating AI content. Still, two considerations from the guidelines are relevant for this.

First, the DSA's focus on generated and manipulated content has a \textit{technology-neutral scope}, as stated in the guidelines. Unlike the AI Act, which focuses only on content generated by AI systems (as defined), the DSA thus covers generated or manipulated content regardless of the techniques used to create this. This scope could include synthetic content produced via CGI-techniques, advanced editing tools (e.g., Photoshop), or other 2D/3D-rendering techniques such as game-footage. We therefore adopt a similar technology-neutral scope, which, we believe, is also methodologically necessary for manual annotation, since distinguishing between advanced generation and manipulation techniques (beyond standard editing) is increasingly impossible. For readability and consistency with platform terminology and prior work, we however keep using \textit{AI} as an overarching term throughout this study. During annotation we also additionally record the likely production technique used for each item, to be able to separately report on content that is likely produced via generative AI techniques. 

Second, the guidelines state that for something to be considered as AI, a generation or manipulation must go \textit{beyond standard edits and minor technical corrections}, such as cropping, colour adjustments, minor technical corrections, compressions, or rescaling. This, in contrast to alterations that change the meaning or substance of content, such as removals, replacements, or insertions of objects, face modifications, or compositions that modify the representation of subjects. 
The guidelines further state that manipulations that build on existing content (whether synthetic or not) still qualifies, and the same goes for mixed human-AI content. 
We adopt these consideration, but extend them to our technology-neutral scope. 

\paragraph{Step 2: What counts as a deepfake?\\} The guidelines parse the deepfake definition into four cumulative legal criteria: (i) resemblance of (ii) existing (iii) persons, objects, places, entities or events (iv) that would falsely appear to a person to be authentic or truthful \cite{AI-Act-guidelines}. We discuss each below in the order in which we apply them during annotation, which follows their logical dependencies. 

The guidelines also highlight the importance of considering the reasonably foreseeable deployment context of the content during assessment. This includes considering the diverse composition of a potential audience, rather than that of an average viewer. As our audit concerns public image and video posts on social media, we expect these to also reach people with lower digital literacy levels. We therefore annotate from the perspective of such a viewer, and thereby adopt a low treshold for each deepfake criteria. Importantly, our unit of analysis is the visual content of a post, whereby accompanying audio only serves as supporting evidence, given the difficulty of assessing audio consistently across languages. 
\begin{itemize}
    \item \textbf{(i) Persons, objects, places, entities or events.} We follow the guidelines' definitions of each term, where objects, for instance, is defined as: ``realistic, inanimate material items, including buildings, artworks, machinery, consumer goods etc.'' 
    \item \textbf{(ii)} \textbf{Existing.} As per the guidelines, it suffices that simulated subjects or events ``resemble someone or something that exists, can plausibly exist or could have plausibly existed in reality''. Content that defies the laws of nature or physics, including lifeforms not accepted in biology (e.g., dragons) generally falls outside this scope. We apply this rule holistically, based on this dominant character of the content rather than on individual elements.
    \item \textbf{(iii) (Appreciably) resemblances.} The guidelines state that content resembles a subject if their is a high degree of similarity between the deepfake and the subject being simulated. The content, however, does not need to be identical, and assessment should be objective and case-by-case. As assessing similarity against a specific real subject is difficult for diverse social media content, we operationalise this criterion through a focus on the content's style: whether the content realistically imitates the characteristics of a genuine recording, relative to the type it appears to be (e.g., camera footage requires a photorealistic style). The guidelines seem to support this reading, as it is noted that a high degree of photorealism makes it likely that the content resembles existing subjects. 
    \item \textbf{(iv) Falsely appears authentic or truthful.} Lastly, the guidelines state that photorealistic image and video content likely constitutes a deepfake, but that this is not yet determinative. Content must also falsely appear as authentic (as to the source or creation process) or truthful (as to its veracity). Assessment hereby does not depend on a creators' intention to deceive, and should consider the content as a whole, including the deployment context, resemblance, and substantive message. Contexts in which audiences expect non-authentic content, such as AI-generated special effects in film productions, are noted as examples where AIGC in that case would likely not be a deepfake. In our audit context, this criterion would, however, likely be satisfied if the first three criteria are met, and fail if any is not. We nevertheless assess it separately to capture exceptions (such as a news item clearly discussing an AI-generated segment). 
\end{itemize}

\paragraph{Step 3: Visual AI disclosures within the content\\} One category of signals we deliberately exclude from our deepfake assessment, namely whether content already contains a clear visible AI disclosure, such as a watermark, sticker, or overlaid text mention (or AI-mentions in the post catpion). This could be particluarly relevant for criterion four. However, taking this into account would conflate our assessment with the labelling outcomes we audit for (which includes any creator-applied labels). We therefore record these signals separately as a third step during annotation. 
The guidelines also do not offer any guidance on how such signals influence a deepfake assessment, as they are primarily written to help determine whether or not to include one. 

\section{RQ2: Platform Practices for AI-Labelling}\label{4}
Having set out the EU's expectations for AIGC labelling, we examine how the four platforms have operationalised them. Section~\ref{4-1} covers the four platform's stated AI-labelling policies, label designs, and public reporting. Section~\ref{4-2} examines  their detection approaches, including through upload experiments with AIGC.  

\begin{table*}
\centering
\begin{threeparttable}
\caption{Overview of the AI-labelling policies of the four platforms as of August 8 2026, including label options, creator disclosure requirements, detection methods, and label design choices.}
\label{tab:platform-ai-labeling}

\footnotesize
\setlength{\tabcolsep}{4pt}
\renewcommand{\arraystretch}{1.15}

\begin{tabularx}{\textwidth}{
  @{} l c L{1.85} c L{0.85} L{0.45} L{0.85} @{}
}
\toprule
\raisebox{0.5\normalbaselineskip}{\textbf{Platform}} &
\shortstack{\textbf{Creator}\\\textbf{label}} &
\multicolumn{1}{c}{\shortstack[c]{\textbf{Creator labelling requirement}\\\textbf{(scope)}}} &
\shortstack{\textbf{Auto-}\\\textbf{labelling}} &
\multicolumn{1}{c}{\shortstack[c]{\textbf{Detection}\\\textbf{methods}}} &
\shortstack[c]{\textbf{Label source}\\\textbf{shown}} &
\multicolumn{1}{c@{}}{\shortstack[c]{\textbf{Label}\\\textbf{visibility}}} \\
\midrule

\textbf{Instagram} &
\checkmark\tnote{a} &
``\textit{photorealistic video or realistic-sounding audio
that has been digitally generated or altered, including
with AI.}'' \cite{instagram_AIlabeling_page} &
\checkmark &
\textbullet~Meta AI tools\newline
\textbullet~Industry-standard AI signals\tnote{c} &
\centering\checkmark\newline
\textit{In post description} &
Some labels directly visible; others in description only. \\

\addlinespace[0.8em]
\textbf{TikTok} &
\checkmark\tnote{b} &
``\textit{AI-generated or significantly edited content
that shows realistic-looking scenes or people}'' \cite{tiktok_integrity_guidelines} &
\checkmark &
\textbullet~TikTok AI effects\newline
\textbullet~C2PA metadata &
\centering\checkmark &
All labels directly visible. \\

\addlinespace[0.8em]
\textbf{X} &
\checkmark &
--- &
\checkmark &
\textit{Not specified} &
\centering--- &
All labels directly visible. \\

\addlinespace[0.8em]
\textbf{YouTube} &
\checkmark &
``\textit{AI to meaningfully alter or generate
photorealistic content}.'' \cite{youtube_main} &
\checkmark &
\textbullet~YouTube AI tools\newline
\textbullet~C2PA metadata\newline
\textbullet~Internal system &
\centering--- &
Some labels directly visible; others in description only. \\

\bottomrule
\end{tabularx}

\begin{tablenotes}[flushleft]
\label{tab:platform-ai-policies}
\scriptsize
\item[]
\quad \textsuperscript{a} + Also an extra ``AI creator'' account-label option.
\quad \textsuperscript{b} May also be via a ``clear caption, watermark, or sticker''.
\quad \textsuperscript{c} Excludes signals indicating `modified with AI'.
\end{tablenotes}

\end{threeparttable}
\end{table*}

\subsection{Platform Policies, Designs and Reporting}\label{4-1}
\paragraph{Platform Policies.}
Most platforms have introduced specific AIGC measures in recent years \cite{gao2026governance}, but as these policies frequently change, we map the current state of labelling-relevant rules, with additional details relevant to our audit context, as shown in Table \ref{tab:platform-ai-policies}.
The table is based on an early August 2026 inspection of platforms' main AIGC policy and information pages and their interfaces   \cite{instagram_AIlabeling_page, meta_labeling_feb2024, instagram_AIlabeling_creators, tiktok_support_ai, tiktok_integrity_guidelines, x_authenticity, x_media_literacy, youtube_main, YouTubeTeam2026AILabels}. 

All four platforms now offer users the option to label their own content, in line with the DSA's expectation for providing such a functionality (as listed in Article 35(1)(k)). X is the only platform without an explicit requirement for creators to use this functionality. X, however, does provide a detailed list of prohibited categories of synthetic and manipulated media \cite{x_authenticity}. The other three platforms do define a user-labelling requirement, but the scope of these does not seem to directly align with the DSA's terminology. Instead, the platforms have operationalised this to a focus on (photo)realism. Instagram's scope is the narrowest, as its creator-labelling requirement does not cover images. TikTok further also allows users to disclose AIGC via a ``clear caption, watermark, or sticker''. 

\paragraph{Label designs.} Platforms' approaches diverge widely in how labels are presented, including in terminology, label types, label options, placement, and visibility. Appendix~\ref{appendix_all_labels} gives a full overview of the different label types as found on the platforms between August 13-20 2026. 

Relevant to the DSA's expectation for ``prominent markings'' is that each platform offers at least one directly visible label, as also shown in Figure~\ref{fig:labels} for the mobile interface. YouTube and Instagram, however, also place AI-labels for some categories of AIGC solely in the post description (see Table \ref{tab:platform-ai-labeling}), which requires one or two clicks to reveal.  Instagram's AI labels were previously reported to not be visible on the web interface \cite{aiforensics-core}, but this is not the case at this time. 

Platforms also differ in what their labels disclose. Labels (or related descriptions) on Instagram, TikTok, and YouTube can indicate the source signal that was used to add the label (e.g., an industry-standard AI signal). TikTok and Instagram thereby also show the distinction between creator- and (automatic) platform-applied labels. YouTube is the only platform that explicitly highlights synthetic audio use (which labels we do not take into account in this study), and X and TikTok also use distinct visible labels to content from their own AI models (Grok and CapCut). Instagram further also recently introduced a new type of AI label: a user-applied account-level creator label \cite{instagram_AIlabeling_creators}. 

\paragraph{Platform reporting on AI-labelling outcomes in the DSA Transparency Database.}  
The DSA requires platforms to inform users of their content moderation decisions through so-called statement of reasons, which they also need to submit to the publicly available DSA Transparency database \cite{dsa_transparency_db}. This database, however, offers limited insight into AI-labelling results. For the four platforms studied here, there is no option to filter specifially for labelling decisions. And while there is a filter for `synthetic media', only Instagram and X seem to use this. Still, only Instagram's data allows a meaningful breakdown of its statement of reasons for synthetic media-tagged content: we find that 9.31\% of the statement of reasons from January 2025 to September 2026 were for content tagged as synthetic media, of which 99,97\% was removed or demoted. 

The absence of comparable data, and the overall lack of data on AI-labelling measures, make it impossible to determine from the Database data how much platforms' identify, label, or remove synthetic media, which provides further motivation for the (black-box) audit of AI-labelling outcomes in Section \ref{5}. 

\subsection{Platforms' AI Detection Approaches}\label{4-2}
\paragraph{Stated methods.} Platforms disclose limited information about how they detect AIGC (Table \ref{tab:platform-ai-labeling}). The methods they name publicly mostly fall into two categories: signals from their own AI tools, and provenance signals from industry standards, such as metadata or C2PA Content Credential signals. YouTube also refers to a recently introduced internal detection system, but without specifying details \cite{YouTubeTeam2026AILabels}, while X does not name any detection methods at all. 


\begingroup
\newcommand{\aYes}{\cellcolor{aigcyes}\ding{51}}
\newcommand{\aDesc}{\cellcolor{aigcdesc}\(\circ\)}
\newcommand{\aNo}{\cellcolor{aigcno}\ding{55}}
\newcommand{\aNA}{\textcolor{black}{N/A}}
\newcommand{\aMeta}{Metadata}
\newcommand{\aCtwo}{C2PA verifiable}

\begin{table*}
\caption{Results of the AIGC upload experiments (21 September 2026).
For files generated using ten generative AI tools, the table reports whether
each platform had automatically added an AI label 24 hours after uploading.
Platforms labelled 36 of the 59 uploads, despite most files containing
machine-readable AI signals.}
\label{fig:upload-tests-2}
\centering

\resizebox{0.85\linewidth}{!}{%
\begin{minipage}{\linewidth}
\centering
\small
\setlength{\tabcolsep}{3pt}
\renewcommand{\arraystretch}{1.28}
\renewcommand{\tabularxcolumn}[1]{m{#1}}

\begin{tabularx}{\linewidth}{@{}
  >{\raggedright\arraybackslash}m{0.14\linewidth}
  >{\raggedright\arraybackslash}m{0.235\linewidth}
  *{7}{>{\centering\arraybackslash}X}
@{}}
\toprule
\multirow{2}{*}{\textbf{AI tool}} &
\multirow{2}{*}{\shortstack[l]{\textbf{Signals embedded}\\\textbf{at generation}}} &
\multicolumn{2}{c}{\textbf{Instagram}} &
\multicolumn{2}{c}{\textbf{TikTok}} &
\multicolumn{2}{c}{\textbf{X}} &
\textbf{YouTube} \\
\cmidrule(lr){3-4}\cmidrule(lr){5-6}
\cmidrule(lr){7-8}\cmidrule(l){9-9}
& &
\shortstack{Video\\(Reels)} & Image &
Video & Image &
Video & Image &
\shortstack{Video\\(Shorts)} \\
\midrule
\textbf{Meta AI} & \aMeta &
\aNA & \aYes & \aNA & \aYes & \aNA & \aNo & \aNA \\
\textbf{CapCut} & \aMeta, \aCtwo\newline (+ visible watermark) &
\aDesc & \aDesc & \aYes & \aYes & \aNo & \aNo & \aYes \\
\textbf{Grok} & \aMeta, \aCtwo &
\aYes & \aYes & \aYes & \aYes & \aYes & \aYes & \aNo \\
\textbf{Gemini} & \aMeta, \aCtwo, SynthID (+ visible watermark) &
\aYes & \aYes & \aYes & \aYes & \aYes & \aYes & \aYes \\
\midrule
\textbf{Adobe} & \aMeta, \aCtwo &
\aDesc & \aDesc & \aYes & \aNo & \aNo & \aNo & \aDesc \\
\textbf{HappyHorse} & \aMeta\newline (+ visible watermark) &
\aDesc & \aNA & \aYes & \aNA & \aNo & \aNA & \aYes \\
\textbf{Kling AI} & \aMeta\newline (+ visible watermark) &
\aNo & \aNo & \aYes & \aNo & \aNo & \aNo & \aNo \\
\textbf{Luma} & Visible watermark only &
\aNo & \aDesc & \aNo & \aNo & \aNo & \aNo & \aYes \\
\textbf{OpenAI}\textsuperscript{a} &
\aMeta, \aCtwo, SynthID, OpenAI verifiable &
\aYes & \aYes & \aNo & \aNo & \aYes & \aYes & \aNo \\
\textbf{Runway} & \aMeta, \aCtwo, SynthID &
\aNA & \aYes & \aNA & \aNo & \aNA & \aYes & \aNA \\
\bottomrule
\end{tabularx}

\par\vspace{5pt}
\raggedright
\footnotesize
\textbf{Key:}\quad
\colorbox{aigcyes}{\strut\ding{51}} AI label added\quad
\colorbox{aigcdesc}{\strut\(\circ\)} AI label in description only\quad
\colorbox{aigcno}{\strut\ding{55}} No AI label added
\par\smallskip
\scriptsize
\textsuperscript{a} OpenAI: ChatGPT for images; OpenAI Playground for video.
\par
\end{minipage}%
}

\end{table*}
\endgroup

\paragraph{Testing platforms' detection methods.} To examine how these (automated) detection approaches function in practice, we tested the platform's detection approaches using uploads of AIGC from ten widely used generative AI models, on 21 September 2026.
For each model, we generated a photorealistic image and video with a clear deep fake-style prompt: ``generate a hyperrealistic [image/video] of a person walking down the street'' (in 9:16 format). Following, we checked whether any standard machine-readable AI signals or visible watermarks were present in the content, using four freely available verification methods: metadata inspection for AI-signals (via \cite{exiftools2026}) the (C2PA) Content Credentials Verify tool \cite{cai2026verify}, OpenAI's image verification tool \cite{openai2026verify}, and a Google SytnhID verification via Google's Gemini model \cite{googlesynthid} (details on the markings found are included Appendix \ref{appendix-upload-experiments}). We then uploaded the 17 generated files (not all models allowed both image and video generations) to each platform as public posts from an empty account using standard upload settings. After 24 hours, we recorded whether the platforms had applied an automatic AI label to the posts, and deleted each post to limit exposure to real users. 

Table~\ref{fig:upload-tests-2} shows the results of this experiment. Platforms only labelled 36 of the 59 uploads (61\%), with 14/17 on Instagram (82\%), 10/17 on TikTok (59\%), 7/17 on X (41\%), and 5/8 on YouTube (63\%) labelled. While this is higher than the results of earlier experiments from \citet{Indicator2026AILabeling}, platforms still miss some standard AI provenance signals, as 15/17 of the generated files carried detectable AI provenance signals. Notably, some platforms did not detect content from other platforms' proprietary models. 

Before deleting the posts, we also downloaded the uploaded content from the platforms to check whether AI provenance signals survived platform processing, and whether any signals about the platform-applied AI labels were attached to the downloaded files. This matters as users might download and re-upload AIGC elsewhere. Downloading of posts via the user interface was, however, only possible on Instagram (for Reels), TikTok (all files), and X (images only). The YouTube posts were downloaded via the creators' environment. 

Using the same four detection methods, we find that most original AI provenance signals seem to be stripped by platforms, apart from the signals that are detectable via OpenAI's and Google's verification tools. TikTok is the only platform that thereby seems to consistently embed signals from its own platform-based AI-labels to the content, which were found both in the content's metadata and via the (C2PA) Content Credentials Verify tool. This is largely not the case for the other platforms, where only YouTube showed C2PA-signals in some of the downloaded posts. This while the DSA electoral guidelines' recommends that labels are retained when content is re-shared on platforms \cite{eu_dsa_electoral_guidelines}. 

Taken together, Section \ref{4} shows that platform practices for AI-labelling diverge substantially across different dimensions, with several practices that seem to diverge from the DSA's expectations and from the state-of-the-art. Next, Section~\ref{5} examines how these practices translate into actual labelling outcomes.

\section{RQ3: AI-Labelling Outcomes}\label{5}
This section addresses RQ3 by measuring what content the four platforms actually label in practice. For this, we use a large-scale dataset of 10,722 recent public posts containing an image or video, collected through identical keyword searches via each platform's standard recommendation settings. We analyse this dataset in two ways: We first manually annotate a 500-post subsample with a codebook that follows the legal operationalisation of Section~\ref{3-3}. This to identify any (AI) generated or manipulated posts, and posts those meet the four deepfake criteria, which can compare against any platform-based AI-labels found. Second, we also measure the overall prevalence of AI labels across the full dataset, including over post upload dates.

The figures presented here must be read with caution. Our dataset and the annotated subsample is small, context-specific, and manual annotation of AIGC is inherently uncertain.
Plus, cross-platform comparisons are complicated by differing rules, user bases, label types, and extractable signals.
Section~\ref{Limitations} gives more details on these limitations.

\subsection{Methods}
\paragraph{Data Collection.} 
We collected recent public posts containing image or video content across the four platforms, using comparable sampling settings. To target content relevant to the DSA's systemic risks, we sampled by keyword, with 15 keywords relating to civic discourse, public security, electoral processes, and/or fundamental rights contexts (e.g., \textit{flood, war, trump}), and five relating explicitly to AI and deepfakes (\textit{ai, genai, deepfake, aigenerated, madewithai}). We included the latter because AIGC frequently also includes content that relates to systemic risks (e.g., AI depictions of politicians). Collection on Instagram and YouTube was limited to Reels and Shorts, as this is most comparable with TikTok's data. On X, we, however, collected both images and videos, as image content often also appeared in a video format on the other platforms. 

Posts were collected at the end of August 2026, using fresh accounts, and scraping each platform with the default `For You' recommendation settings, which surfaces content that is relevant, still available, and actively recommended to users. We used several different libraries and browser automation tools for scraping, including TikTok-Content-Scaper~\cite{bukold2025tiktok} and SOAP~\cite{SOAP}. In our case, requesting access to a research API (per DSA Article 40(12)) was not an option since platforms severely restrict the metadata provided per post and do not include AI-labelling signals, such as platform-added AIGC labels \cite{API-Audit-Luka}. Post searches were keyword-based, but hashtags were used when recommendations ran out; only on TikTok, we primarily relied on hashtag-based search, but the recommendations surfaced both keyword- and hashtag-related matches. AI keyword posts were largely collected last, as to limit their influence on recommendations surfaced for the other keywords. 

We collected 150 posts per keyword, amounting to 3,000 posts per platform. We then filtered this dataset to include only posts uploaded after 25 August 2023, when the DSA's systemic risk rules took effect. We also removed duplicates, and multi-content posts on Instagram (carousel posts) and X (reposts with multiple media files in them), which would be impractical to assess in manual annotation. This gives a final dataset of 10,722 posts. Appendix~\ref{appendix data collection} reports all the keywords used and detailed dataset statistics. 

\paragraph{Recording of AI-labels.}
As each platform uses different types of AI labels (see Section \ref{4}), the extraction of AI-label signals differed by platform. For \textbf{Instagram}, the scraped metadata distinguished between creator self-disclosed labels, platform-classifier labels, and account-level AI-creator signals, as well as whether a label is directly visible or only shown in the post description. We could, however, not identify when content was both creator- and platform-tagged. In \textbf{TikTok}'s metadata, we could also see whether signals were creator-, platform-applied, or both. For \textbf{X}, we extracted the two AI labels (the general AI and the Grok label) from each post's HTML code, since the scraped metadata did not include label signals. For \textbf{YouTube}, we similarly collected label signals from the HTML code, which allowed us to distinguish between directly visible labels and labels added only in the post description. YouTube does not visibly distinguish creator- versus platform-applied labels, but we were able to extract a provenance mention that was sometimes included in the disclosure description (e.g., ``Info from OpenAI"). For YouTube, we also excluded the signals that indicated the presence of generated audio (tracks) in the content.  

We did not analyse caption-based AI disclosures at this stage, as many captions discuss AI without showing AIGC, and we believe such disclosures do not meet the same standard as platform-based AI labels, as they are not always visible during casual scrolling, are less clear then standardly used labels (particularly also for non-English viewers). 

\paragraph{Manual annotation.}
Annotation followed the three-step process set out in Section~\ref{3-3}, and the full annotation codebook is included in Appendix~\ref{appendix:codebook}. 
In step 1, we assign a primary label that indicates whether the visual content is generated or manipulated (beyond standard edits), using four coding options: \textit{fully generated or manipulated}, \textit{partially generated or manipulated }(e.g., video's mixing generated and human-created segments, or partial image manipulations), \textit{no signs of generation or manipulation}, or \textit{unclear}. We adopt an deliberate conservative approach for the primary label and only code content as generated or manipulated when there are clear (visual) signals for this. Examples for such signals are given in the codebook, hereby partly drawing on the AI-detection guide of \citet{aiforensics-core}. For content coded as generated and manipulated, we also separately record the likely production technique used, either \textit{generative AI,} \textit{other 3D-rendering techniques} (e.g., game-based footage), \textit{digital graphic content} (e.g., digital illustrations or infographics), or \textit{other}. We record digital graphic content separately, as for such content it is increasingly impossible to tell whether it was created by generative AI, or other methods. For content that seems to mix production techniques, we give priority to an AI classification.  

Next, \textbf{step 2} assesses the four deepfake criteria for the generated or manipulated content posts. Following our interpretation of EU's legal guidance, annotators code the deepfake criteria from the perspective of a viewer with a lower digital literacy level, based on a single viewing of the content at normal speed. 

In \textbf{step 3}, we record whether any AI disclosures are present in the visual content, such as visible AI watermarks, or AI-overlaid texts, because such signals could count as a type of AI-label/disclosure. We record these signals separately, as such signals might not carry equal weight to standard platform labels. 

Annotation was carried out by three authors, each with substantial experience working with AIGC and social media data. We first refined the codebook on data outside the final dataset, which took several iterations, as the EU's legal guidelines often did not offer concrete guidance for the wide diversity of social media content. In this process, we developed several guiding decision rules for the codebook, for instance on how to handle content with multiple distinct AI-generated segments, for which we adopted a holistic evaluation rule with particular attention to the opening section of the content, before users scroll away.

We then extracted a random 500-post subsample (roughly 5\%), stratified by platform (125 each) and keyword type (100 risk-keyword posts and 25 AI-keyword posts per platform). All three coders independently annotated the first 50 posts, using a custom-built annotation tool with the codebook rules built-in (as shown in \ref{appendix-annotation-tool}, and provided on GitHub). Coders reached Krippendorff's alphas of 0.51-0.91 across the main labels (74-98\% agreement), and disagreements were discussed on the basis of the codebook. Following, the remaining 450 posts were double-coded in varying pairs, with Cohen’s Kappas of 0.59-0.69 for the main labels (or 79-98\% agreement), indicating moderate to substantial agreement between coders \cite{cohenkappa}. All disagreements were resolved through discussion, in some cases supported by additional internet searches and account inspections. Appendix~\ref{appendix-annotation-statistics} reports the full agreement statistics. We publicly release the post metadata and annotations of our dataset also on GitHub, but not the media content itself, in line with EU copyright and privacy rules.

\subsection{Results}
\paragraph{Results for expert-annotated subsample (N = 500).}
\begin{figure*}[t!]
    \centering
    \includegraphics[width=0.85\textwidth]{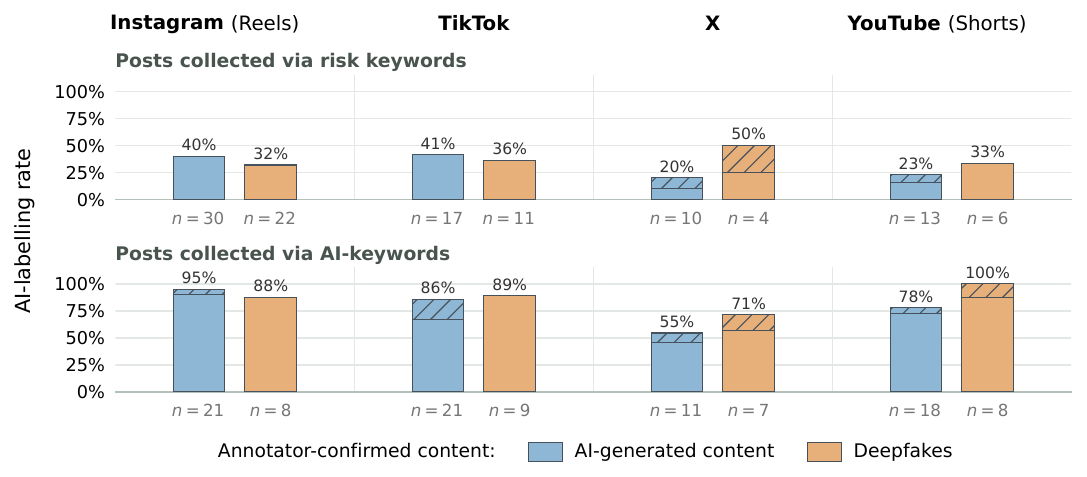}
    \caption{\textbf{AI-labelling rates for expert-confirmed AI-generated content (AIGC) and deepfakes} in the 500-post annotated subset. Solid segments (\solidkey{}) indicate standard platform-based AI labels found; striped segments (\stripedkey{}) indicate in-content visual AI disclosure (e.g., a visible watermark) found in the absence of a platform-based label. The data is split by platform and the keyword category used to collect the posts. The subset includes 100 risk-keyword collected posts and 25 AI-keyword collected posts per platform.
    AIGC in this figure includes both AI-generated and 3D-rendered content. 
    }
    \label{fig:main-figure}
\end{figure*}
Within the 500 annotated posts, we identified 177 posts as fully or partially generated or manipulated. Of these, 75 were found to be deepfakes, as per our interpretation of the EU's legal criteria; 43 deepfakes were found in the 400 posts collected via the systemic risk related keywords (11\%), the other 32 deepfakes (or: 32\%) were found in the 100 AI-keyword posts. Almost all deepfakes were identified as likely produced with generative AI techniques (72/75). The remaining three used other 3D-rendering techniques, or were coded as digital graphic content, which -- as per our codebook -- could have been created by any tool.

The generated or manipulated content that was not coded as a deepfakes, failed the deepfake criteria in two main ways: Around one third of these posts was coded as digital graphic content, which includes digital infographics or illustrations. These largely failed all deep fake criteria. The remainder, which were created via generative AI or other 3D-rendering techniques, mostly failed criterion (ii) of (\textit{existing}) and criterion (iii) of (\textit{resemblance}).

Most importantly, Figure \ref{fig:main-figure} relates these findings to the AI labels that were found on the platforms. We find that only a third (14 out of 43, or 33\%) of expert-identified deepfakes in systemic risk contexts carried a platform-based AI label that discloses its artificial origin to users. Figure \ref{fig:main-figure} further breaks down these findings by platform and by keyword types used to find the post (systemic risk-related keywords or AI-related keywords). This as AI-keyword posts might carry less risks due to the explicit AI-mention in the caption (although such keywords are often not directly visible during casual scrolling). The Figure also splits the labelling rates between deepfakes and a broader category of AIGC, which includes all content identified as AI-generated or created via 3D-rendering techniques. We, however, exclude digital graphic content from this Figure, as these could have been produced by any technique, and rarely met any of the deepfake criteria. 

Two patterns stand out. First, labelling rates are generally much higher across platforms for AI-keyword posts, which could be because AI-keywords influence platforms' detection approaches, or that creators more often label such posts (which we cannot determine with certainty due to the lack of such signals), or that creators are less likely to actively remove provenance signals from such uploads. Second, in the highest-risk category, of AIGC and deepfakes found via systemic risk keywords, labelling rates range between 20 to 50\% across the four platforms, which shows that at least half of such posts are left unlabelled. These labelling rates are, however, based on low amounts of posts per platform.

We also recorded whether any other visible AI-markings are present in the visual content itself, such as provider watermarks or overlaid AI mentions, which is particularly relevant given the AI Act's requirement for \textit{deployers} to visibly disclose deepfakes. We find 31 of the 141 AI-generated or 3D-rendered posts carrying such a signal. Such signals might however not always be clear to an average user, as 14 of these posts for example only carried a small Gemini sign in the bottom-right corner. Encouragingly, 25 of these 31 posts also already carried a platform label and/or a AI-mention in the post caption. We further found only one creator AI-keyword mention in the post captions of deepfakes that were gathered via the systemic risk related keywords.

The deepfakes posts in this sample further attracted significant user engagement, with a median of 160,000 views and an average of 8 million views (calculated for the 59/71 of posts with view data available). We also assessed whether any deepfake posts where deleted (approximately 2-4 weeks after initial collection), and found one  deep fake to be deleted on Instagram (which was labelled), three on TikTok (of which one was labelled), and none on X or YouTube. Four deepfakes that were found on X did, however, have an age-restriction signal. Each of these was sexual content. Apart from such realistic sexual content, the annotation also surfaced other highly concerning AIGC, including fabricated war footage, deepfake segments `hidden' within larger compilation videos, and numerous AI depictions of politicians.

Finally, the figures of AIGC found likely reflect a lower bound, as we exclude posts marked as `unclear' (14 posts), and because our conservative labelling approach required clear signals for content to qualify as generated or manipulated. This means the most sophisticated deepfake content may have been missed by us, which is also somewhat reflected in the data, as five posts that were coded as `unclear' or `no signs of generation or manipulation' did in fact include a platform AI-label. For each of these posts, coders spent significant time analysing them, but did not find clear enough signals. 


\paragraph{Results on the full dataset (N = 10,722).} 
We also measure AI-label prevalence across the full dataset, while noting the absence of ground truth data for the true share of AIGC per platform. Table \ref{tab:table-full-dataset} reports the main statistics. We find a notable share of AI-labelled posts across platforms, also in posts scraped via systemic risk keywords, where an average 9\% of such posts carries AI label across platforms. We find further that roughly 40\% of AI labels is added by creators, but such data is only available for Instagram and TikTok. We also checked what share of AI labels is directly visible to users (as opposed to only being visible in post description), and found that this is the case 60\% of the AI labels on Instagram and 82\% on YouTube.

\begin{figure}[h]
    \centering
    \includegraphics[width=1\linewidth]{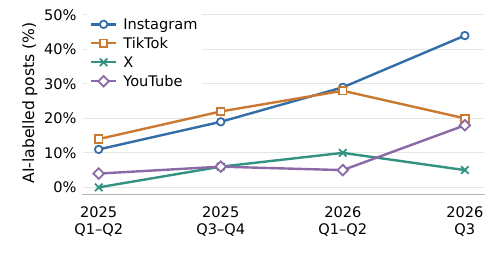}
    \caption{AI-labelled posts as a share of the total amount of posts uploaded on each platform per time period, as found in the full dataset (N = 10,722).}
    \label{fig:over-time}
\end{figure}

Figure \ref{fig:over-time} further presents the share of AI-labelled posts across post upload periods. This shows two main things: First, platforms also label (or have retroactively labelled) posts from older periods, which is relevant because our data collection shows that such posts are still surfaced by the recommendation algorithms. Second, label shares rise steadily across post upload times, which may reflect a growing volume of AIGC uploads, improved platform detection, more creator-labelling, or a combination of these factors. Notably, YouTube shows a sharp increase in labelling rates for 2026 Q3 uploads, which might relate to its May 2026 roll-out of a new internal AI detection system \cite{YouTubeTeam2026AILabels}.

\begin{table*}[h]
\caption{The left-side columns show the share of AI-labelled posts in the full dataset (N = 10,722), split by platform and by the keyword category used to collect the posts. The right-side columns show, among the AI-labelled posts, the share of labels that are directly visible, and the share that was creator-applied.}
\label{tab:table-full-dataset}
\centering
\small
\renewcommand{\arraystretch}{1.2}
\setlength{\tabcolsep}{10pt}

\begin{tabular}{@{}lccc@{\hspace{24pt}}cc@{}}
\toprule
& \multicolumn{3}{c}{%
  \shortstack{\textbf{AI-labelled posts}\\
  \footnotesize (\% of posts in each category)}}
& \multicolumn{2}{c}{%
  \shortstack{\textbf{Label characteristics}\\
  \footnotesize (\% of AI labels)}} \\
\addlinespace[2pt]
\textbf{Platform}
& All posts & AI keywords & Risk keywords
& Visible directly & Creator-labelled \\
\midrule
\textbf{Instagram} & 34\% & 74\% & 19\% & 60\%  & 42\% \\
\textbf{TikTok}    & 21\% & 50\% & 11\% & 100\% & 37\% \\
\textbf{X}         &  5\% & 18\% &  1\% & 100\% & Unknown \\
\textbf{YouTube}   & 12\% & 33\% &  5\% & 82\%  & Unknown \\
\bottomrule
\end{tabular}
\end{table*}

\section{Limitations}\label{Limitations}
This study does not aim to assess legal compliance. This is for two main reasons. First, we cannot fully observe platform's internal practices (such as invisible metadata, their detection systems used, what content is deleted, or some AI-labelling signals), or their ongoing exchanges with regulators. Second, the DSA's expectations for AI-labelling (and their scope) are still open to interpretation. Our empirical results on AI-labelling outcomes thus rests on the European Commission's current operationalisation of these expectations (Section \ref{3}), and our findings concern observable outcomes on AI-labelling measures and the feasibility of measuring them. 

Results should also be read in light of our specific audit context and methods used. This includes the time period of the experiments: reported results are a snapshot based on August-September data, and platform policies, label designs, and detection methods change frequently, as our own results also show. And also relates to our dataset, keywords used, and scraping configurations. We aimed to gather posts from DSA-relevant systemic risk contexts, using keyword-based sampling, but these may not always directly relate to systemic risks. Plus, keyword-based sampling may also have influenced the platforms recommendations' systems and thereby affected the AI-labelling prevalence numbers (though not whether a given post met the deepfake criteria). Our dataset further is small relative to the daily amounts of posts uploaded on these platforms, and specific to this topic, the reported figures can thus not be interpreted as platform-wide rates. The results from the upload experiments are similarly based on a small dataset, and uploads were done from a single account in a short time period, which may have affected the results. 

Direct platform-comparisons on the basis of the reported labelling rates should also be interpreted with caution, as platforms differ in what content is uploaded, what content their recommendation systems surface, and in the types of AI label signals that can be seen and extracted. Plus, this study does not take into account the content removal strategies that platforms also use to handle AIGC and deepfakes. 

Finally, manual annotation of AIGC is inherently uncertain. Our labelling approach was deliberately conservative for the primary label, as we required clear signals before coding something as (AI) generated or manipulated, which means we might have missed the most sophisticated deep fakes, and that the figures for the primary label likely reflect a lower bound. We further had moderate agreement on some of the four deep fake criteria during annotation (prior to resolving all differences through discussion), which was in part due to annotation errors, but also indicates the difficulties of consistently applying the EU's guidelines (as operationalised in our codebook) to highly diverse social media content.

\section{Discussion and Implications for the DSA and Platform Policies (and Beyond)}\label{7}
\paragraph{Labelling measures and uptake have advanced.} Across the four platforms, we find that AI-labelling is now broadly established, and observe clear improvements from earlier audit studies: All four platforms now offer creators the option to add AI labels and also apply labels automatically. 
Labelling rates for AIGC in our annotated subsample are higher than those reported in earlier studies (although direct comparisons are difficult due to varying methods, see Section \ref{2}). 
We also find a larger share of labels applied (automatically) by platforms, rather than via creator disclosures. This may reflect improvements in platform detection, and possibly a wider adoption of machine-readable marking signals embedded at generation, as mandated by the AI Act. Our upload experiments show that these measures can work well for platform transparency when signals are left intact, and platforms consistently check for them, which a successfully implementation of the AI Act's requirements can hopefully further contribute to. 

\paragraph{Labelling rates remain relatively low, especially for AI and deepfake content found risky contexts.} 
In relative terms, labelling remains partial. Of all AIGC (AI-generated or 3D-rendered content) identified in our annotated sample, 58\% carried a platform-based AI label or other visual disclosure in the content. However, when looking at AIGC and deepfake posts found via systemic risk related keywords, only 20-50\% carried such an AI disclosure across the four platforms. Labelling rates were notably higher for AIGC and deepfakes collected via the AI-related keywords. A possible explanation is that creators that use an AI keyword in their caption are also more likely to self-disclose (which we cannot reliable measure in the data across platforms), or less prone to strip detectable signals from the content. AI captions could also have influenced platform's detection systems. 

\paragraph{Platforms' labelling implementations still diverge widely.}
Platforms' creator labelling requirements differ in scope and do not always seem to align with the DSA's expectations. Instagram, for instance, excludes images, and X does not impose any labelling rules for creators. Labels also differ in terminology, placement, types used, and visibility. These differences could in part be explained by the limited DSA-specific guidance that exists, and the difficulties in operationalising the DSA's expectations. But they also present a clear opportunity to improve on both the uptake and the effects of AI labels. Users and creators would likely benefit from aligning the rules, terminologies, and label design and types (as would auditors examining these practices). DSA guidance and communication could address these gaps by, for instance, recommending the use of the standard EU AI icons, as included in the AI Act's Code of Practice for Article 50 \cite{EU2026TransparencyCodeOfPractice}.

\paragraph{Other opportunities for improving the uptake and effect of AI labels.} Our audit also points to several other readily-available opportunities that could help to improve the uptake and effect of AI labels. These include: showing all AI labels directly in the visual content instead of sometimes only in post descriptions (as TikTok and X do); highlighting the differences between creator-and platform applied labels (as TikTok and Instagram do); more reliably and consistently detecting standard AI-provenance signals; retaining invisible AI provenance signals after uploading, and also attaching such signals when a platform-based AI label is added; increasing and standardising public disclosures on labelling outcomes in the DSA transparency database; and sharing more detailed signals to external auditors. These recommendations demonstrably fall within the state-of-the-art, as other platforms already implement them, and also highlight the potential for DSA policy makers to help in sharing best practices. 

\paragraph{The difficulties of reliable detecting and labelling AIGC and deepfakes on platforms.}
Still, these suggestions do not take away from the genuine difficulties that platforms also face in reliably detecting and labelling AIGC and (risky) deepfakes. 
When AI provenance signals are stripped from content, reliable automatic detection becomes very complex. Other (ML-based) detection approaches would likely introduce difficult challenges in handling mislabelling, for which no guidance yet exists. A potential solution might, however, be found on YouTube, which has introduced a new internal detection system, but thereby also offer creators the option to remove automated labels applied by this system (but not when applied based on more reliable signals) \cite{YouTubeTeam2026AILabels}. 
Furthermore, platforms address the risks of AIGC and deepfakes aso through other enforcement measures, including post deletions, visibility restrictions, and age restrictions, which we have not examined in this study. 

Beyond automated efforts, our annotation also highlights difficulties for the manual detection of increasingly realistic AIGC and deepfakes. Substantial efforts were required in many cases to find clear enough signals to classify content as AIGC. Plus, the EU's legal focus on deepfakes created an additional coding challenge, as operationalising and applying this focus across highly diverse social media content proved difficult (as also reflected in our inter-annotator agreement statistics). Even when relying on detailed legal guidelines, where to draw the line between for deepfakes remained highly subjective, particularly for partial AIGC (mixed content) and multi-segment videos. However, as these EU's guidelines were not developed for the DSA, dedicated DSA guidance could potentially bridge this gap, and thereby draw inspiration from our codebook. However, this focus on deepfakes does not always directly map to risks and harms on social media, as we also observed plenty content not passing the deepfake criteria, but which could still be harmful (e.g., clearly stylised depictions of politicians). 

\section{Conclusion}
AI labels on social media aim to provide transparency about AIGC and deepfakes to users, and thereby aid in reducing various potential risks (including deception, misinformation, and manipulation), and overall help preserve the integrity of the online information ecosystem. Our findings show that platform labelling measures (albeit imperfect) are already delivering such transparency at scale. While labelling rates remain partial, their absolute volume on the most popular platforms is already substantial, which, apart from transparency, also allows for curation efforts for \textit{AI slop}. Increasingly such transparency also comes from automatic detection rather than from creator disclosures (alone).

Still, AI-labelling is not a complete answer to the risks that AIGC poses on social media. We find that labelling rates remain relatively low on some platforms and for AIGC and deepfakes found in the most risky contexts, where transparency matters most. Prior work also highlights how the use of AI transparency measures can carry risks of its own, as for example unlabelled AIGC can be perceived as more truthful. We also note the difficulty of the task for platforms, particularly when acting against sophisticated bad actors who remove AI provenance signals, and acknowledge the responsibility for users and creators to contribute. At the same time, we identify several readily available improvements for platforms that could likely help increase the uptake and clarity of AI labels. Most importantly, we recommend that platforms, with the help of policy-makers, focus on aligning their AI-labelling rules, designs, types, and detection approaches.

\paragraph{Future work.} The complexities observed in our work highlight several potential areas for future work. Audit studies so far have largely focused on only a few of the most popular platforms, future work could broaden this scope to identify other emerging best practices, including for AI-generated audio and text detection measures. Future work could also examine whether AI transparency obligations from other jurisdictions might help to further define or operationalise the scope of platform's AI-labelling efforts under the DSA. Finally, another direction could be to examine whether multi-content LLM-based (agentic) classifiers could be used to reliably scale the assessment of deepfakes under EU law. To support these efforts, we release our methods, tooling, and datasets on GitHub \textbf{(link added upon publication)}. 

\section*{Acknowledgements}
Bram Rijsbosch and Konrad Kollnig are supported by the RegTech4AI AiNed Fellowship Grant, provided via the Dutch National Growth Fund (NGF) under file number NGF.1607.22.028. Luka Bekavac and Henry Tari are researchers under the CoCoDa project, which funded by SNSF (Swiss National Science Foundation) under grant number 10004598. The authors further wish to thank the other members of the RegTech4AI and CoCoDa projects, in particular Kamil Szostak and Lucas G. Uberti-Bona Marin, for helpful discussions and comments. 

\paragraph{Generative AI Use.} The authors acknowledge the use of AI-assistants (Claude, Gemini) for minor text editing, and to help improve the clarity of the author-written text. The content and intellectual contributions are those of the human authors. The authors also acknowledge the use of AI-coding tools (Claude and Gemini) that supported coding, data scraping, data analysis, and the creation of figures.  

\bibliography{aaai2026}
\newpage
\appendix
\begin{center}
    \huge\bfseries Appendices
\end{center}
\vspace{1em} 
\section{Data Collection and Dataset Overview}
\label{appendix data collection}
\paragraph{Publicly shared data.}
All data (including annotations) that can be publicly shared (in light of EU rules, such as for privacy) will be uploaded to GitHub [\textit{upon publication}]. \\

\paragraph{Data collection.}
Table \ref{tab:keywords-sampling} provides an overview of the keywords that were used to scrape the posts on each platforms (150 per keywords). The DSA-relevant systemic risk keywords relate specifically to risks around civic discourse, public security, electoral processess and fundamental rights .

\begin{table}[h!]
\centering
\small
\renewcommand{\arraystretch}{1.2}
\begin{tabular}{p{3.2cm} | p{4.3cm}}
\toprule
\textbf{Category} & \textbf{Keywords} \\
\hline
\midrule
\textbf{AI-explicit keywords} &
\textit{ai}, \textit{genai}, \textit{madewithai}, \textit{aigenerated}, \textit{deepfake} \\[4pt]
\hline
\textbf{DSA-systemic risk relevant keywords} &
\textit{kids}, \textit{disaster}, \textit{war}, \textit{iran}, \textit{woman}, \textit{explosion}, \textit{putin}, \textit{trump}, \textit{police}, \textit{fight}, \textit{storm}, \textit{celebrity}, \textit{migrants}, \textit{politician}, \textit{climate}\\
\end{tabular}
\caption{AI-explicit and DSA systemic risk-relevant keywords used in sampling.}
\label{tab:keywords-sampling}
\end{table}

\begin{figure}[h]
    \centering
    \includegraphics[width=1\linewidth]{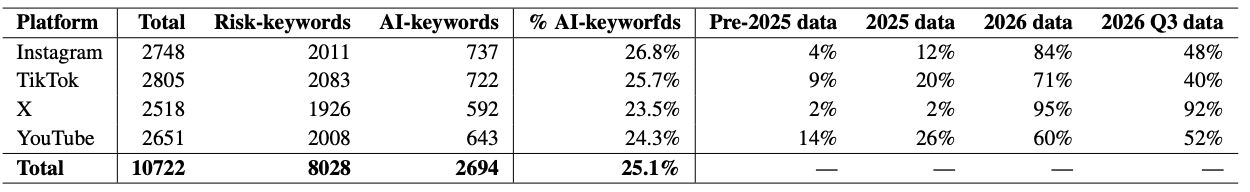}
    \caption{Posts in the dataset, including keyword distribution and temporal breakdown percentages across platforms.}
    \label{fig:data-appendix-overview}
\end{figure}

\paragraph{Data overview.}
Table \ref{fig:data-appendix-overview} provides an overview of the amounts of posts included in the dataset for each platform, with details on keyword distributions.  

\section{Annotation Statistics}\label{appendix-annotation-statistics}
\begin{figure}[h!]
    \centering
    \includegraphics[width=0.85\linewidth]{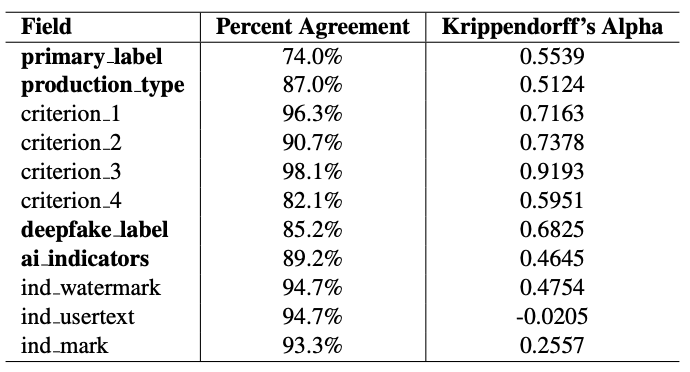}
    \caption{Inter-annotator agreement statistics for the initial 50-post subset (3 coders).}
    \label{annotator-1}
\end{figure}

Table \ref{annotator-1} shows the statistics for each of the coded categories for the 3-coder annotated subset of the first 50 posts. 

\begin{figure}[h!]
    \centering
    \includegraphics[width=0.85\linewidth]{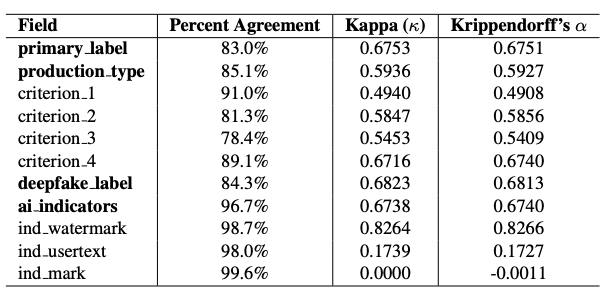}
    \caption{Inter-annotator agreement statistics for the 453-post subset (2 coders).}
    \label{fig:annotator-2}
\end{figure}

Following, Table \ref{fig:annotator-2} shows the same statistics, but then for the double-coded 453 posts subset that followed. We coded 3 extra posts due to sampling mistakes that included pre-2023 data in the first 50 posts, these were removed from the analyses later.

\begin{figure}[h!]
    \centering
    \includegraphics[width=1\linewidth]{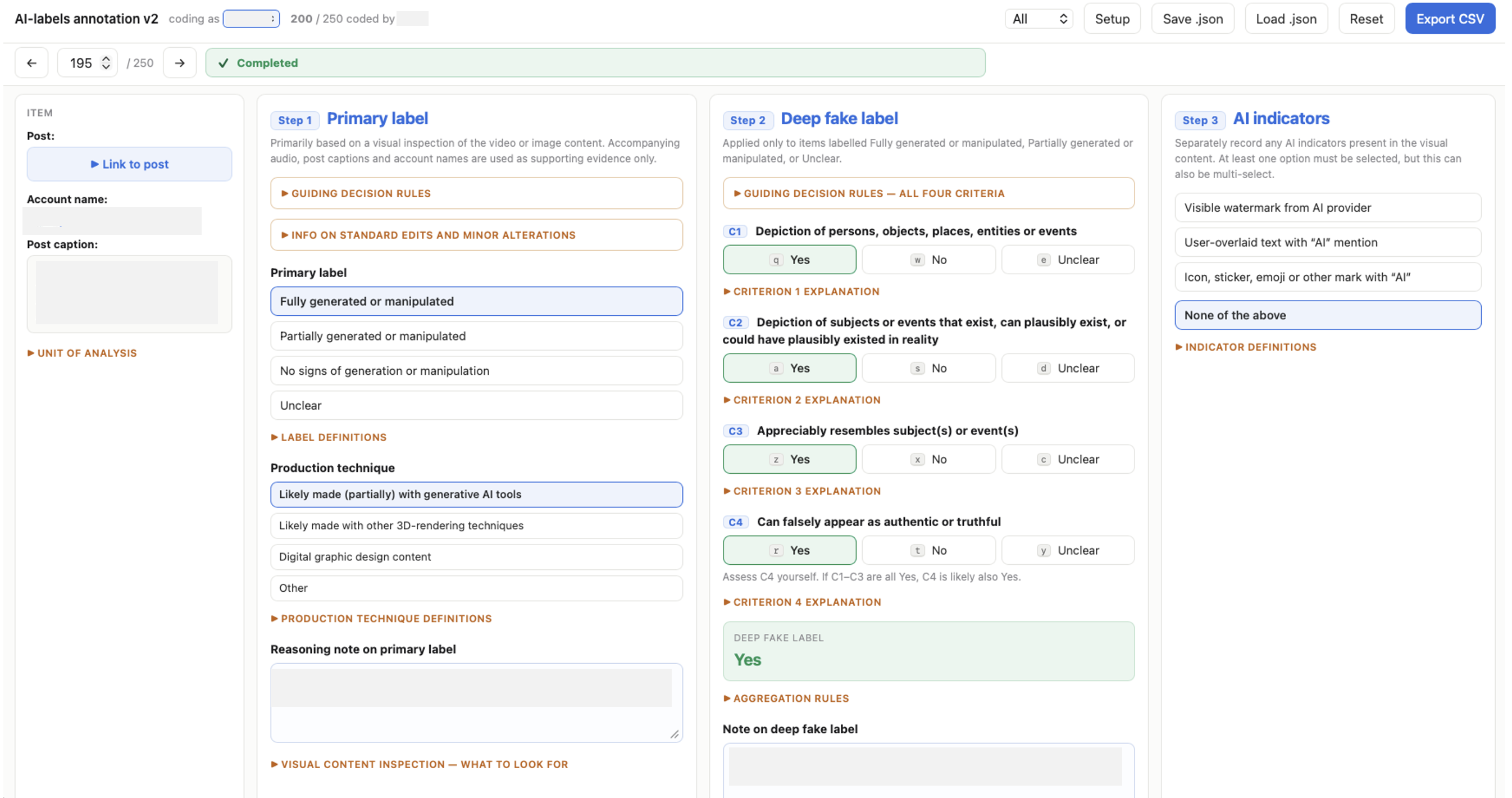}
    \caption{Screenshot of the custom-built annotation tool used by the coders}
    \label{fig:annotaiton-tool}
\end{figure}

\paragraph{Annotation Tool}\label{appendix-annotation-tool}
We add also a screenshot of the custom-built annotation that was used by all three annotators for the coding (Figure \ref{fig:annotaiton-tool}. This tool is also uploaded to GitHub, and includes the criteria, details and guiding examples/decision rules from the codebook. 

\newpage
\section{Upload Experiments Dataset}\label{appendix-upload-experiments}
\begin{figure*}[h!]
    \centering
    \includegraphics[width=0.95\textwidth]{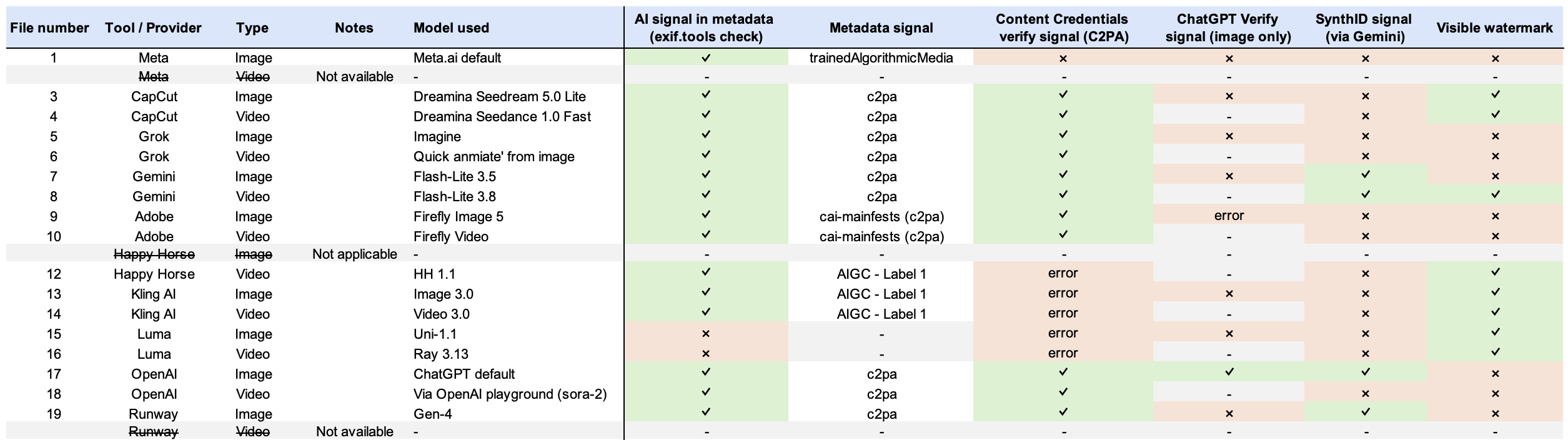}
    \caption{An overview of the dataset that was for the upload experiments, including the AI signals found in each of the files after generation, which can include different types of machine-readable marking signals and/or visible watermarks.}
    \label{fig:upload-experiment-data}
\end{figure*}
Figure \ref{fig:upload-experiment-data} shows the dataset that was used for the upload experiments. This Figure highlights the different invisible or visible markings that were found in each generated files prior to uploading. The dataset can be found on the GitHub page linked to the project. Each of these files was then uploaded to the four social media platforms (with the exception of YouTube that does not allow image uploads). Following, 24h's after upload, we recorded whether any (automatic) AI-labels were added by the platforms, after which we downloaded and deleted the posts. 

\section{Overview of AI Label Types Found on Each Platform}\label{appendix_all_labels}
\begin{figure*}[h!]
    \centering
    \includegraphics[width=0.95\textwidth]{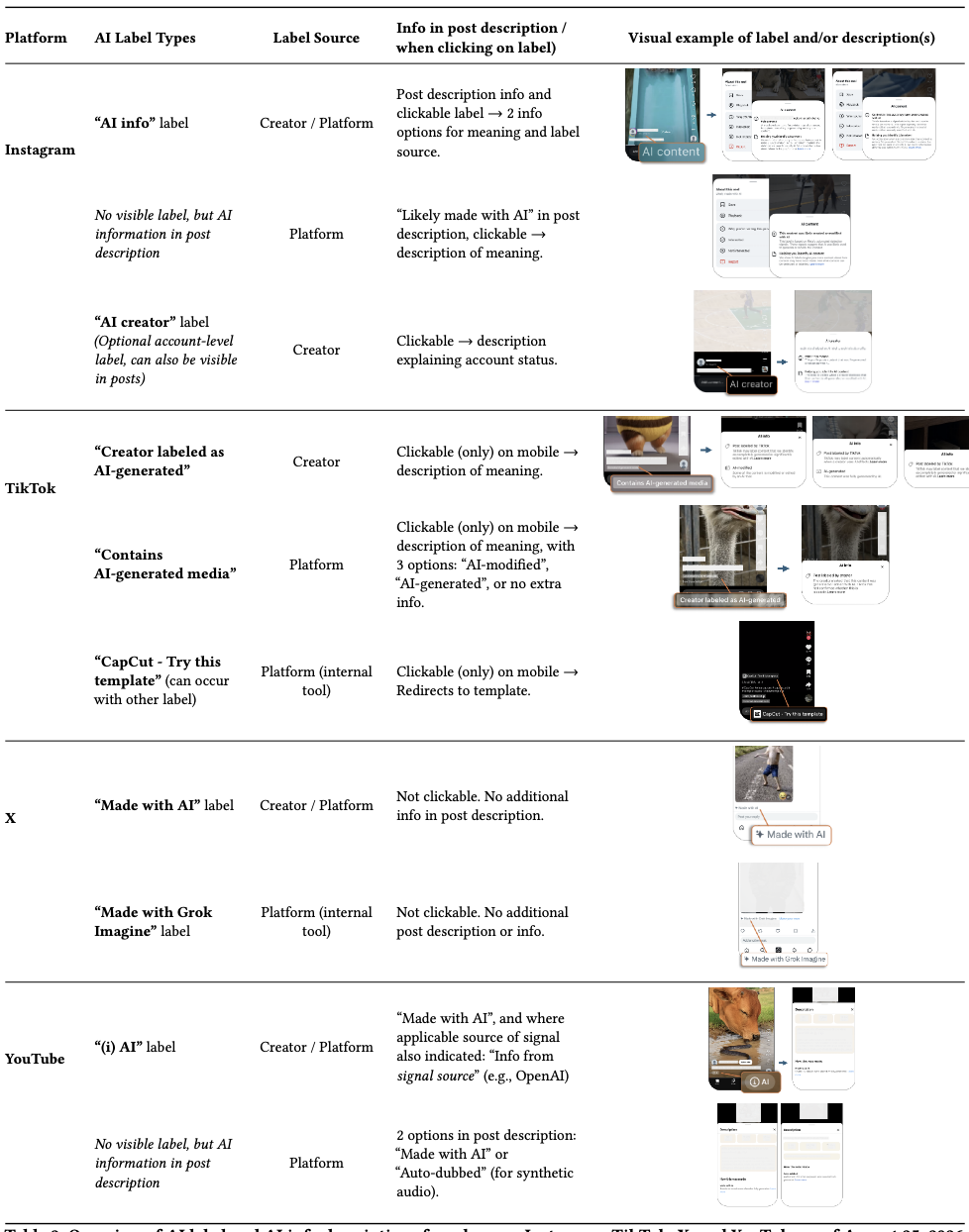}
    \caption{Overview of AI label and AI info descriptions found across Instagram, TikTok, X, and YouTube as of August 25, 2026. The figure columns show screenshots of public posts with the AI labels and the info in the post descriptions.}
    \label{fig:all-labels-appendix}
\end{figure*}

Figure \ref{fig:all-labels-appendix} on the next page provides a full overview of the different types of AI labels and related post descriptions that are identified in the four platforms of this audit (as of August 25, 2026). The AI labels look similar on mobile vs. web interfaces, but the placement of the label, and the options to click on it or not (as indicated) can differ per interface.
\clearpage

\section{Annotation Codebook}\label{appendix:codebook}
\subsection{General Procedure and Recorded Fields}
\textbf{Unit of Analysis:} A public social media post that includes an image or video.

\begin{itemize}
\item
  \textit{Multi-content (carousel) posts:} are excluded from the sample.
\item
  \textit{Audio}: annotation is based on the visual elements in the images and videos. Accompanying audio only serves to highlight potential visual anomalies or to prompt deeper inspections.
\end{itemize}

\noindent\textbf{Procedure:} Annotation proceeds in three steps:

\begin{itemize}
    \item \textbf{Step 1:} Assign a primary content label: \emph{Fully generated or manipulated}, \emph{Partially generated or manipulated}, \emph{No signs of generation or manipulation}, or \emph{Unclear.}
    \subitem [IF Step 1 = \emph{Fully generated or manipulated}, \emph{Partially generated or manipulated}, or \emph{Unclear} → separately record the likely production technique used for the generated or manipulated elements.]
    \item \textbf{Step 2:} IF Step 1 = \emph{Fully generated or manipulated}, \emph{Partially generated or manipulated} or \emph{Unclear} → Assess the four criteria for the \textbf{Deepfake} label.
    \item \textbf{Step 3:} separately record if any AI-indicator signals are present within the visual content.    
\end{itemize}

\noindent\textbf{Final Recorded Fields}

\begin{itemize}
\item
  \textbf{Primary label:} \emph{Fully generated or manipulated}
  \textbar{} \emph{Partially generated or manipulated} \textbar{}
  \emph{No signs of generation or manipulation} \textbar{}
  \emph{Unclear}
\item
  \textbf{Production technique:} \emph{Likely made with generative AI
  tools} \textbar{} \emph{Likely made with other 3D-image rendering
  techniques} \textbar{} \emph{Graphic design content} \textbar{}
  \emph{Other}
\item
  \textbf{Reasoning note on primary label:} \emph{{[}text{]}}
\item
  \textbf{Criterion 1:} \emph{Yes} \textbar{} \emph{No} \textbar{}
  \emph{Unclear}
\item
  \textbf{Criterion 2:} \emph{Yes} \textbar{} \emph{No} \textbar{}
  \emph{Unclear}
\item
  \textbf{Criterion 3:} \emph{Yes} \textbar{} \emph{No} \textbar{}
  \emph{Unclear}
\item
  \textbf{Deepfake label:} \emph{Yes} \textbar{} \emph{No} \textbar{}
  \emph{Unclear}
\item
  \textbf{Reasoning note on deepfake label:} \emph{{[}text{]}}
\item
  \textbf{AI indicator:} \emph{Visible watermark from AI provider}
  {[}\emph{+ provider selected}{]} \textbar{} \emph{User-overlaid text
  with `AI' mention} \textbar{} \emph{Icon, sticker, emoji or other mark
  with `AI'} \textbar{} None of the above\\
  (at least one item must selected, but this can also be multi-select)
\end{itemize}
\noindent\textbf{----\/-\/-\/-\/-\/-\/-\/-\/-\/-\/-\/-\/-\/-\/-\/-\/-\/-\/-\/-\/-\/-\/-\/-\/-\/-\/-\/-\/-\/-\/-\/-\/-\/-\/-\/-\/-\/-\/-\/-\/-\/-\/-\/-\/-\/-\/-\/-\/-\/-\/-\/-\/-\/-\/-\/-\/-\/-\/-\/-\/-\/-\/-\/-\/-\/-\/-\/-\/-}
\subsection{Step 1: Primary Label}
\textbf{Procedure for Primary Label}
\begin{itemize}
\item
  Annotation is based on the visual elements of the video or image content.
\item
  Signals from audio or post context (captions and account names) can be
  used as well, but solely as supporting evidence. The primary signals
  to pay attention to during annotation are outlined below, and build on
  the inspection guide from
  \citet{aiforensics-detection-guide}.
\item
  Videos are in principle watched in full to avoid missing any
  AI-generated segments, but skipping is permitted when the
  classification seems sufficiently clear.
\item
  Record a short reasoning about the assessment.
\end{itemize}

\noindent\textbf{Generation/manipulation signals}

\textit{Primary signals (in visual content)}
\begin{itemize}
\item
  \emph{\textbf{Inspect for content violations:}} such as in
  physics/biology, inconsistent proportions, unnatural elements,
  disappearing or missing elements (e.g., shadows), wrong text
  inscriptions, or unmatching elements (e.g., sound-visuals).
\item
  \emph{\textbf{Inspect for style and format indicators:}} such as
  AI-stylised glow, game-footage indicators, vanishing points/shadows.
\item
  \emph{\textbf{Inspect for generation indicators:}} such as AI
  watermarks, icons, stickers, texts with AI, any indications of
  watermark removals, or signs of other generator tools (such as gaming
  visuals) (Note: separately record any AI-generation indicators
  also in step 3.)
\end{itemize}

\textit{Supporting evidence signals:}

\begin{itemize}
\item
  \emph{\textbf{Audio:}} accompanying audio can help to highlight visual
  anomalies, such as unnatural lip movements, mouth blurring, mechanical
  jaw motions, or lip-syncing that clearly fails to match the spoken
  syllables. Clearly synthetic, robotic, or highly suspicious
  voice-overs can also act as a trigger for further scrutinisations. But
  purely audio-based manipulations can \textbf{not} qualify a post as
  `generated or manipulated'.
\item
  \emph{\textbf{Length of the video segments:}} longer continuous video
  segments are more unlikely to have been generated via state-of-the-art
  AI methods.
\item
  \emph{\textbf{Post caption and account name:}} inspect for any
  mentions of generation or manipulation, such as hashtags with AI
  keywords, mentions of tool names, or other textual disclosures.
  \emph{(Note: non-English post captions are translated prior to
  annotation using Google Translate)}.
\end{itemize}

\noindent\textbf{Guiding Decision Rules}

\begin{itemize}
\item
  \textbf{Conservative approach:} label conservatively, and choose the
  fully/partially- generated or manipulated labels only if there are
  clear signals for such a classification.
\item
  \textbf{Technology-neutral focus:} content can be generated or
  manipulated by generative AI systems, or by any other image/video
  generation and manipulation techniques, such as CGI, photo editing
  tools (like Photoshop), 2D/3D-image rendering techniques, or other
  graphic design methods.

  \begin{itemize}
  \item
    \textbf{Digital graphic design content:} digital graphic design
    content (such as digital illustrations or infographics) is always
    labelled as generated or manipulated content, as it is increasingly
    difficult to tell whether such content was made with AI tools, other
    generation methods, or by manual efforts (for example through copy
    and pasting in powerpoint). We separately record for such content in
    the `production technique' label.
  \end{itemize}
\item
  \textbf{Exception for standard editing and minor alterations}:
  standard content edits or minor alterations are not counted as
  generations or manipulations. Whether that is the case, however,
  requires a case-specific assessment. The following can be seen
  examples of standard edits and minor alterations, as outlined in the
  AI Act guidelines \cite{AI-Act-guidelines}:

  \begin{itemize}
  \item
    Format conversions, technical compressions, noise reduction or
    removal for enhanced clarity, without changing the meaning or the
    substance of the content;
  \item
    Minor cropping, minor colour adjustments or corrections, lightening
    or darkening, sharpening or enhanced clarity or other technical
    corrections;
  \item
    Removal of dust spots caused by a dirty lens, removal of red-eyes,
    deleting and obscuring backgrounds that are visible in the original
    file, pixelation or blurring of faces;
  \item
    Rescaling of a video clip, dynamic range compression and
    equalisation; limited video stabilisation;
  \item
    Standard adjustments to colour and contrast, minor adjustments to
    playback speed, minor corrections to level the horizon of an image,
    applying pixel filters to amplify certain parts of an image, or
    applying colour maps to grayscale images, edge image completion,
    converting a black \& white to colour image or video and vice
    versus;
  \item
    Pixel filling for aspect ratio adaptation or enhanced clarity,
    automatic transition clips, other non-substantive edits for cosmetic
    and technical purposes.\\
  \end{itemize}
  
Editing goes beyond standard editing if the content is changed in a
material way (substantive modifications, structural changes etc.) that affect its meaning, style or intent. The following are examples of non-standard edits:

\begin{itemize}
\item
  Removal, replacement or insertion of objects or persons in existing
  images and videos that changes meaning and substance of the content;
  face replacement or substantial facial modification.
\item
  Generation of realistic video depicting events that did not occur;
  altering the body shape or the skin colour of a person; extreme
  lightening, darkening, colour and contrast changes that change the
  meaning, intent and messaging of the content
\item
  Creation of composite images or video clips that modifies the
  representation of persons, objects, events or facts, any other
  substantial alteration of the content.
\end{itemize}
\end{itemize}

\begin{itemize}
\item
  \textbf{Other exclusions from the annotation decision:} the following aspects should not be taken into account when assessing the primary label and production technique:

  \begin{itemize}
  \item
    \textbf{Content quality aspects:} do not consider the realism or
    quality of the generated/manipulated content \emph{(as} t\emph{his
    is part of Step 2.})
  \item
    \textbf{User-added overlays:} disregard any simple user-added
    overlays, such as overlaid texts, emojis, or stickers.
  \item
    \textbf{Reposts or reupload frames}: disregard the repost/reupload
    frame that can be shown, and label solely based on the content shown
    within this frame.
  \item
    \textbf{Screen recordings of synthetic content:} for content that
    contains screenshots or screen-recordings of generated/manipulated
    content, label based on the content shown, and not the capture of
    it.
  \item
    \textbf{Intro and outros}: disregard standard intro and/or outro
    graphics, such as news logo graphic visuals, as these do not
    influence the main content displayed.
  \end{itemize}
\end{itemize}

\noindent\textbf{The Labels}\\

\textbf{Fully generated or manipulated: }There are clear signs that the visual content consists entirely of generated imagery, or the manipulation affects the whole scene (e.g., a whole-scene manipulation of real footage, or a full style transfer).\\

\textbf{Partially generated or manipulated:} There are clear signs that the visual content combines generated/manipulated elements and non-generated/manipulated (human-created) elements, beyond simple user-added overlays and minor/standard editing.\\
This category includes mixed content (e.g., non-generated videos with short AI-generated segments or images in it, or generated background scenes overlaid with a real person speaking), and manipulated content where manipulations only affect part of the content (e.g., face swaps, insertions, or removals of objects/persons).\\
The proportion or duration of the generated or manipulated elements does not influence the classification; a small or short generated/manipulated segment still suffices, provided that it goes beyond the exceptions listed above (such as those for standard editing and minor technical corrections).\\

\textbf{No signs of generation or manipulation:} There are no signs or signals that the visual content was generated or manipulated.\\

\textbf{Unclear:} Unclear is assigned when there are some signs or signals that the content includes generated or manipulated content, but these signs are not strong enough to reliably classify it as such. Unclear is also assigned when the production format makes classification impossible. Unclear is not an annotation failure, but a valid outcome.\\

\noindent\textbf{Flag for production technique used}\\
\noindent Note: The production technique is only annotated for items labelled as \textit{Fully generated or manipulated, Partially generated or manipulated, or Unclear.}\\

\noindent Separately flag the likely production technique used for the generated or manipulated content, which is similarly done based on any evidence from the visual inspection, or supporting evidence from the audio and post caption. Select from the following options:\\
\begin{itemize}
\item
  \textbf{Likely made (partially) with generative AI tools:} content
  created, altered, or synthesized using generative AI models.
\item
  \textbf{Likely made with other 3D-rendering techniques:} this can
  include content such as game-based footage, flight/machinery
  simulations, virtual reality content, architectural walkthroughs,
  3D-animated movies, or non AI-based CGI-footage in movies.
\item
  \textbf{Digital graphic design content:} this can include content such as digital illustrations, digital infographics/charts or pricing tables, graphic designs of logos and brand marks, screenshots of a graphics designed app environment, or slide-style graphics. 
  \item \textbf{Other}\\
\end{itemize}

\noindent\textbf{Guiding decision rules:}
\begin{itemize}
\item Digital graphic design content is labelled separately, even though it could have been created via AI or 3D-rendering techniques.
\item For generated/manipulated content that seems to combine multiple different production techniques: an AI flag takes priority over the other techniques, and 3D-rendering similarly takes priority over digital graphic design content.
\end{itemize}

\noindent\textbf{----\/-\/-\/-\/-\/-\/-\/-\/-\/-\/-\/-\/-\/-\/-\/-\/-\/-\/-\/-\/-\/-\/-\/-\/-\/-\/-\/-\/-\/-\/-\/-\/-\/-\/-\/-\/-\/-\/-\/-\/-\/-\/-\/-\/-\/-\/-\/-\/-\/-\/-\/-\/-\/-\/-\/-\/-\/-\/-\/-\/-\/-\/-\/-\/-\/-}

\subsection{Step 2: Deepfake Label}
\textit{Note:} The  label is applied only to items labelled as \emph{Fully generated or manipulated}, \emph{Partially generated or manipulated}, or \emph{Unclear}.\\

\noindent\textbf{Procedure for Deepfake Label\\
}Assess the four criteria that together determine the deepfake label, based on the decision rules and examples outlined below: \\

\noindent\textit{Aggregation Rules:} For the first three criteria, record: \textit{Yes}, \textit{No}, or \textit{Unclear}.\\
IF the first three criteria are all marked either \textit{Yes} or
\textit{Unclear} → assess the fourth manually, and again record: \textit{Yes}, \textit{No}, or \textit{Unclear.} \\
IF the first three criteria include a \textit{No →} the fourth criterion
is automatically assigned a \textit{No}.\\

\noindent\textit{The final Deepfake Label is then determined as follows:}
\begin{itemize}
\item
  \textbf{Yes:} If all criteria are marked \emph{Yes}.
\item
  \textbf{No:} One or more criteria are marked \emph{No}.
\item
  \textbf{Unclear:} One or more criteria are marked \emph{Unclear}, and
  no criteria are marked \emph{No}.\\
\end{itemize}

\noindent\textbf{Guiding Decision Rules}

\begin{itemize}
\item
  \textbf{Annotation persona perspective:} assessment of the four
  criteria should not be based on a hypothetical ``average'' person
  expected to be exposed to the content, but should take into account
  the possible diverse composition of a reasonably foreseeable audience,
  with due consideration especially to content that may be perceived by
  persons with lower digital literacy or general knowledge levels
  \cite{AI-Act-guidelines}.\\
  Annotation should therefore be based on the perspective of a
  person with lower digital literacy or technical knowledge, as they
  can generally be expected to be active on social media, and seeing
  publicly available posts. In practice, this means that assessment is
  based on a casual, single watch of a video at normal speed, and flaws
  that can only be discovered through closer scrutiny, pausing, looping,
  or similar checks must also be ignored. Plus, crucially, while the
  primary label requires a conservative approach, this secondary
  assessment thus requires the opposite approach: where in doubt, adopt
  a `yes'.
\item
  \textbf{Partially generated or manipulated content:} For content that
  mixes generated/manipulated and non-synthetic elements, the first
  three criteria are assessed solely based on the
  generated/manipulated elements of the content; Criterion 4 does
  evaluate the generated/manipulated elements in relation to the entire
  content.
\item
  \textbf{Holistic evaluation rule:} assessment of the first three
  criteria is based on a holistic evaluation of the (generated or
  manipulated) content that focuses on its dominant or overall
  character, and not on whether there is a single item or segment
  depicted in the content that could satisfy a criterion, where the rest
  of the content does not (e.g., for criterion 2: a realistic car placed
  in a an overall impossible event such as it driving on a rainbow is
  still a `No'). Particular weight should, however, be given to the
  start of a video, where viewers are highly vulnerable to being misled
  before scrolling away..
\item
  \textbf{Exclusion of any visible AI disclosures from content
  analysis:} Assessment does not take into account any visual
  generation/manipulation disclosures within the content, and is also
  \textbf{not} based on any disclosures present in the post caption
  \emph{(as this is the outcome we audit for.})\\
\end{itemize}

\noindent\textbf{The 4 Label Criteria}\\

\noindent\textbf{Criterion 1: Depiction of persons, objects, places, entities or
events}\\
\noindent\textit{(Options: Yes / No / Unclear)}\\

\noindent The generated or manipulated content depicts subjects (persons, places, objects, or entities) or events (scenes and situations), where subjects and events be described as follows
(following the AI Act guidelines \cite{AI-Act-guidelines}):

\begin{itemize}
\item
  \emph{\textbf{Persons}}: understood as realistic, human beings
  (including digital replicas of real persons, realistic AI-generated
  human avatars or personas, and personal characteristics or
  expressions, such as image, voice, behaviour, performances etc.).
\item
  \emph{\textbf{Places}}: is to be understood as realistic locations.
\item
  \emph{\textbf{Objects}}: is to be understood as realistic, inanimate
  material items, including buildings, artworks, machinery, consumer
  goods etc
\item
  \emph{\textbf{Entities}}: is to be understood as realistic, non-human
  but animate beings including animals or other biological lifeforms.
\item
  \emph{\textbf{Events}}:is to be understood as realistic scenes or
  situations that can involve persons, objects, places and entities
  (e.g. evoking historical events or the depiction of professional or
  consumer services).
\end{itemize}

\emph{Importantly: the quality of the generated or manipulated content,
or the quality of resemblance to the subject or event is \textbf{not}
evaluated in criterion 1.}\\

\noindent\textbf{Criterion 2: Depiction of subjects or events that exist, can plausibly exist, or could have plausibly existed in reality}\\
\noindent\textit{(Options: Yes / No / Unclear)}\\

\noindent The generated or manipulated content depicts simulated subjects
(persons, places, objects, or entities) or events (scenes and
situations) \textbf{that exist, can plausibly exist, or could have
plausibly existed in reality.}\\
Highly improbable, staged, or fictional depictions still qualify. By
contrast, simulated persons, objects, places, entities or events that,
for example, defy the laws of nature or physics or depict lifeforms that
are not commonly accepted in biology (such as e.g. humans flying without
mechanical aids, dragons, or elephants driving cars) and have no
potential to mislead are considered unrealistic and therefore fall
outside the scope.

\emph{\textbf{Guiding Examples:}}

\begin{itemize}
\item
  \emph{Yes:} A depiction of: footballers in a stadium, a politician
  speaking, a deceased, fictional or celebrity person dancing, a natural
  disaster event, a synthetic avatar of a CEO speaking, or an image of a
  product in an advertisement.
\item
  \emph{No:} A depiction of: a sphinx flying around, a person flying
  without mechanical aid, mice arguing in human language.\\
\end{itemize}

\noindent\textbf{Criterion 3: Appreciably resembles subject(s) or event(s)}\\
\noindent\textit{(Options: Yes / No / Unclear)}

The generated or manipulated content appreciably resembles a subject if
there is a high level of similarity between the deepfake content and
the subject (including any of its recognisable elements) or event being
simulated by the deepfake. The content does not need to be identical to
the subject or event. Whether the level of resemblance is appreciable is
a case-by-case assessment, based on, among others, the extent to which
characteristic or distinctive features are represented by the deepfake.

\emph{\textbf{Extra guiding decision rule:}}

\begin{itemize}
\item
  \textbf{Photorealism:} a high degree of photorealism renders it likely
  that the content resembles existing subjects or events \emph{(as noted
  in the AI Act guidelines \cite{AI-Act-guidelines})}.
\end{itemize}

\emph{\textbf{Guiding} \textbf{Examples:}}

\begin{itemize}
\item
  \emph{Yes:} A photorealistic depiction of a politician, a nature
  scene, or a news bulletin. Photorealistic footage of a war from a
  video game.
\item
  \emph{No:} Cartoonish/illustrated depiction of a real-world scene,
  cartoon depiction of a historical event, clear video game graphics,
  highly stylised depiction of a politician.\\
\end{itemize}

\noindent\textbf{Criterion 4: Can falsely appear as authentic or truthful}\\
\noindent\textit{(Options: Yes / No / Unclear)}\\

\noindent ``Whether content `falsely appears to a person to be authentic or
truthful' should be assessed as a whole, taking into account the level
of resemblance, the potential substantive message of the content, the
intended and foreseeable deployment contexts, the environment in which
the content is presented, and the intended and reasonably foreseeable
audience composition and their expectations. However, this assessment is
objective and does not require the intention of the deployer to deceive
or mislead the natural persons exposed to the content for it to
constitute a deepfake.'' \cite{AI-Act-guidelines}.

\begin{itemize}
\item
  \textbf{\emph{``}Content authenticity} refers to whether the content
  is genuinely what it purports to be in terms of its source or creation
  process (including e.g. the involvement of real human beings or
  animals and their actual behaviour or actions, the actual appearance
  or use of objects or the delivery of services, the accurate unfolding
  or course of an event).'' \cite{AI-Act-guidelines}
\item
  ``\textbf{Truthfulness} pertains to the veracity of the content (e.g.
  factual correctness of the representations in the deepfake).''
  \textbackslash \cite{AI-Act-guidelines}
\end{itemize}

\emph{\textbf{Extra guiding decision rules:}}

\begin{itemize}
\item
  IF all three of the first criteria are \emph{Yes}, it is likely that
  the fourth criterion is also a \emph{Yes,} given the deployment
  context (public social media posts) and reasonably foreseeable
  audience (which likely includes people with lower digital literacy
  levels).
\item
  IF any of the first three criterion is \emph{No} the fourth criterion
  is automatically marked as a \emph{No (since content failing criterion
  1 cannot by its definition be a deepfake any more, and content
  failing criteria 2 or 3 cannot appear as realistic or authentic
  anymore).}
\item
  Any AI indicators in the content's visuals and/or post caption are
  again not taken into account in the assessment of criterion 4
  (\emph{as this is the outcome we audit for, and as this is also
  separately annotated in Step 3}).
\end{itemize}

\emph{\textbf{Guiding examples:}}

\begin{itemize}
\item
  For content from certain production or fictional contexts where
  generated or manipulated elements are clearly expected, and where they
  do not target a real performance, the content could \textbf{not}
  falsely appear as authentic or truthful, even if it is in fact
  non-authentic or untruthful content.\\
  \emph{(e.g., a clear movie scene depiction that contains generated
  special effects, or photorealistic game-based footage where there are
  clear visual indicators for it being gaming footage)}
\item
  Similarly, for partially generated or manipulated content where the
  generated elements only play an evidently illustrative or aesthetic
  role, the content may \textbf{not} falsely appear as authentic or
  truthful, even if it is in fact non-authentic or untruthful content\\
  \emph{(e.g., a real speaker in front of a clearly generic generated
  backdrop, a news segment or an educational video discussing and
  showing (AI-)generated footage and clearly depicting it as an exhibit,
  or a real product in an advertisement with a generated background that
  is unlikely to mislead about the advertised product).}\\
\end{itemize}
\noindent\textbf{----\/-\/-\/-\/-\/-\/-\/-\/-\/-\/-\/-\/-\/-\/-\/-\/-\/-\/-\/-\/-\/-\/-\/-\/-\/-\/-\/-\/-\/-\/-\/-\/-\/-\/-\/-\/-\/-\/-\/-\/-\/-\/-\/-\/-\/-\/-\/-\/-\/-\/-\/-\/-\/-\/-\/-\/-\/-\/-\/-\/-\/-\/-\/-\/-}

\subsection{Step 3: AI indicators in the visual
content}\label{step-3-ai-indicators-in-the-visual-content}

Separately record if any AI indicators present in the visual content (True/False). Using the following options:

\begin{itemize}
\item
  \emph{\textbf{Visible watermark from generative AI provider.}}

  \begin{itemize}
  \item
    Separately select a provider from a list of known watermarks, or
    choose: \emph{`Other',} and specify the name of the provider.

    \begin{itemize}
    \item
      \emph{Google} (\emph{Gemini sign), Google (`Veo' text), Kling,
      Seedance/Seedream, Meta (`Meta AI' logo), Runway (`Runway' text),
      Sora (`Sora' watermark,) CapCut (logo).}
    \item
      \emph{Other {[}+ name of provider{]}.}
    \end{itemize}
  \end{itemize}
\item
  \emph{\textbf{Icon, sticker, emoji or other visible marking with `AI'
  in it}} (such as the EU AI icon).
\item
  \emph{\textbf{User overlaid text with `AI'} \textbf{mentioned.}}
\item
  \textbf{\emph{None of the above.}}\\
\end{itemize}

\noindent\textbf{Guiding decision rules:}

\begin{itemize}
\item
  For content that clearly has different generated/manipulated elements
  or segments (such as several AI images depicted after each other),
  each of these elements should have a visible watermark to mark as
  \emph{True}.
\item
  User overlays or icons/stickers/emojis that do not have `AI' in it,
  but that do include other AI-related text mentions, such as tool
  names, are not counted (\emph{as these might not be clear enough
  indicators for the reasonably foreseeable audience that includes
  people with lower digital literacy levels})
\end{itemize}

\end{document}